\documentclass[10pt,aps,prd,twocolumn,nofootinbib,noshowkeys,noshowpacs,longbibliography,
superscriptaddress,tightenlines,floatfix]{revtex4-2}
\usepackage{float,ulem}
\usepackage{graphicx,epsfig,epstopdf,amsmath,amssymb} 
\usepackage{xcolor}
\usepackage[colorlinks=true,linktocpage=true,linkcolor=blue,citecolor=blue,citecolor=red,urlcolor=blue]{hyperref}
\usepackage{orcidlink}
\usepackage{caption}
\usepackage{subcaption}

\newcommand{\be}{\begin{equation}}
\newcommand{\ee}{\end{equation}}
\newcommand{\ba}{\begin{eqnarray}}
\newcommand{\ea}{\end{eqnarray}}
\newcommand{\nn}{\nonumber\\}

\usepackage{xcolor}
\usepackage{ulem}
\definecolor{purple}{rgb}{0.8,0,0.6}

\begin{document}
%************************
%************************
\title{Necessary conditions for causality from linearized stability at ultra-high boosts}
%************************
%************************
\author{Shuvayu Roy\orcidlink{0000-0002-5725-5712}}
\email{shuvayu.roy@iitgn.ac.in}
\affiliation{Indian Institute of Technology, Gandhinagar, Gujarat 382055, India.}
\author{Sukanya Mitra\orcidlink{0000-0002-2401-957X}}
\email{sukanya.mitra10@gmail.com}
\affiliation{School of Physical Sciences, National Institute of Science Education and Research, An OCC of Homi Bhabha National Institute, Jatni-752050, India.}
\author{Rajeev Singh\orcidlink{0000-0001-5855-4039}}
\email{rajeev.singh@e-uvt.ro}
\affiliation{Department of Physics, West University of Timisoara, Bulevardul Vasile P\^arvan 4, Timisoara 300223, Romania.}
%************************
%************************
\begin{abstract}
In this work, we provide a novel method to constrain the causal parameter space of a relativistic hydrodynamic system exclusively from its linear stability analysis at non-zero momenta. Our approach exploits the Lorentz-invariant stability property of causal theories. In boosted frames, the dispersion relation exhibits a feature that we call ``$\gamma$-suppression,'' whereby the higher-order terms in the wavenumber expansion are increasingly suppressed beyond leading order at large boosts. As a consequence, at near-luminal values of Lorentz boost, stability criteria at the spatially homogeneous limit are sufficient to identify the region of the parameter space that satisfies the necessary conditions of causality, even at non-zero momenta. After presenting the general hydrodynamic framework, we test the method in conformal Müller-Israel-Stewart theory and show that it provides an efficient way of deriving the necessary conditions of causality while remaining within the low-energy regime of hydrodynamic validity.
\end{abstract}
%************************
%************************
\maketitle
%************************
%************************
\section{Introduction}
\label{sec:intro}
%************************
%************************
To describe collective systems that can be coarse-grained from an underlying microscopic formulation, effective field theories have proven useful across a range of physical systems. Relativistic hydrodynamics~\cite{Landau1987Fluid} is one such effective theory that finds applications ranging from high-energy physics (facilitated by heavy-ion experiments)~\cite{Baier:2007ix,Huovinen:2006jp,Heinz:2013th,Gale:2013da,Braun-Munzinger:2015hba} to astrophysical problems involving neutron stars~\cite{Andersson:2020phh} and cosmological dark matter~\cite{Cabral-Rosetti:2001tbx}. It is a low-energy, long-wavelength limit of its microscopic origin, i.e., an ``infrared approximation" considering that the high-energy parts of the original solutions are small and do not include fast-growing modes. The physical acceptance of such a theory relies on several benchmark criteria. Some of the major ones state that the governing equations must obey (i) sub-luminal signal propagation, i.e., causality conditions, and (ii) perturbations around equilibrium being bounded over time (not growing indefinitely), i.e., stability criteria~\cite{Toll:1956cya,Hiscock:1983zz,Hiscock:1985zz,Olson:1990rzl}. 

Relativistic hydrodynamic theory is generally formulated via systematic derivative corrections to include dissipative phenomena over the ideal fluid theory (leading term)~\cite{Romatschke:2009im,Kovtun:2012rj}. 
Due to the introduction of dissipation, gradient-corrected theories of relativistic fluids encounter pathologies~\cite{Hiscock:1987zz,Bemfica:2017wps}. The two conditions mentioned above can be used to impose the required constraints of physical applicability over a possibly large number of candidate theories.
This is the reason that the analysis of causality and stability constraints for a given theory in terms of its parameter space (set by the transport coefficients of the underlying theory) is crucial and has become a growing field of current studies~\cite{Denicol:2008ha,Pu:2009fj,Brito:2020nou,Bemfica:2019knx,Bemfica:2020zjp,Kovtun:2019hdm,Hoult:2020eho,Biswas:2022cla,Biswas:2022hiv}. Furthermore, the interconnectedness between these two apparently distinct features has led to fascinating recent studies in the context of the frame-invariant properties of a relativistic dissipative hydrodynamic theory~\cite{Heller:2022ejw,Gavassino:2021owo,Gavassino:2023myj,Gavassino:2023mad,Gavassino:2021kjm,Wang:2023csj,Hoult:2023clg,Hoult:2024cyx}. 

Traditionally, the analysis of causality within the scope of a linearized theory (in the sense of its perturbations) belongs to the physics of its large momenta (large wavenumber)~\cite{Hoult:2023clg}. In this context, the asymptotic causality condition is widely used to check the causal validity of a hydrodynamic theory, which states that the group velocity of a propagating mode at the infinite wavenumber limit must not exceed the speed of light. Several studies~\cite{Fox:1969us,Kundt,Pu:2009fj} argue that the asymptotic group velocity being sub-luminal preserves the causality of a theory as a whole, even if it diverges for some intermediate values of momenta (as long as it is restricted to a finite range), since the Fourier transform of a linear signal has significant contributions from large wavenumbers ($k$). The issue with this analysis is that this large-momentum treatment is well outside the domain of validity for hydrodynamics as a low-energy effective theory. Hydrodynamic gradient expansion, being a divergent series with a zero radius of convergence~\cite{Romatschke:2017ejr,Heller:2013fn,Heller:2015dha} has little ability to reproduce the large-momentum behavior of the underlying UV (ultra-violet) complete theory. Our current work attempts to derive the causal parameter space of a given theory without leaving the low-momentum regime of relativistic hydrodynamics.

To explore the causal parameter space of a relativistic hydrodynamic theory in the low wavenumber limit, we use the property of frame-invariant stability of a theory. The linearized hydrodynamic theory is known to face the paradox that small perturbations around equilibrium could lead to stable solutions in one reference frame but become unstable in another. Recently, it has been shown that this stability-invariance issue has profound consequences for causality, which is rigorously studied in Refs.~\cite{Gavassino:2021owo,Gavassino:2023myj,Hoult:2023clg,Bhattacharyya:2024ohn}. The agreement on stability between Lorentz-boosted observers depends on how the non-hydrodynamic modes are introduced to restore causality~\cite{Basar:2024qxd,Bhambure:2024axa,Gavassino:2025hwz}. In a nutshell, causality is a necessary condition for covariant stability to hold and violating causality results in disagreement between different observers on whether a dissipative system is stable or not.

Following this line of argument, in this work, we utilize the covariantly stable parameter space to identify the causal domain of the theory without leaving the small $k$-limit. 
In Ref.~\cite{Roy:2023apk}, we have already introduced this alternate way of deriving the causal parameter space at the spatially homogeneous limit ($k\rightarrow 0$) of the theory. There, we analyzed the stability-invariance of the dispersion modes (frequencies) at the leading order in $k$. This means that, in any arbitrary boosted frame, if a dispersion mode $\omega$ is expressed by the following expansion in terms of momenta $k$,
\be
\omega=\sum_{n=0}^{\infty}a_n k^n~, \quad a_n={\text{constant coefficients}},
\label{disp1}
\ee
only the leading term $a_0$ has been analyzed there to derive the causal parameter space at $k\rightarrow 0$ limit. The inclusion of the non-vanishing wavenumber in the stability analysis of Fourier modes is known to be quite non-trivial (especially in boosted frames), even if we restrict ourselves to small values of $k$ and truncate the dispersion series~\eqref{disp1} to a certain low order in $k$. In continuation with our recent work~\cite{Bhattacharyya:2025hjs}, we show here that in an arbitrary moving frame with boost velocity $v$, the $k$ expanded dispersion series exhibits a behavior termed ``$\gamma$-suppression", which significantly improves the situation. This can be expressed as the following, 
\be
\omega=\sum_{n=0}^{\infty}a_n \left\{\frac{k}{\gamma}\right\}^n, \quad \gamma=\frac{1}{\sqrt{1-v^2}}~.
\label{disp2}
\ee
In a reference frame moving with near-luminal boost velocity ($v\rightarrow 1$), this suppression becomes sufficiently strong to make the next-to-leading order terms in~\eqref{disp2} practically insignificant. So, in near-luminal Lorentz frames, the mode analysis and, consequently, the stability constraints reduce to that of the spatially homogeneous limit ($k\rightarrow 0$), where the treatment of only the leading term suffices for the analysis. 
We show for the causal Müller-Israel-Stewart (MIS) theory~\cite{Israel:1976tn,Israel:1979wp,Muller:1967zza} that at the $v\rightarrow 1$ limit, the stable parameter space remains the same as for $k\rightarrow 0$, irrespective of the non-zero values of $k$ or the order of $k$ considered in the dispersion series~\eqref{disp2}. Now, since for any causal theory, for every $k$, there must be a common region of parameter space that is stable at all $v$ (including $v\rightarrow 1$), we argue here that the knowledge of the near-luminal stable modes only at the $k\rightarrow 0$ limit can provide the parameter space of the theory that satisfies the necessary criteria for causality, even for the non-zero momenta that include all next-to-leading order terms in the $\omega$ expansion.       

The manuscript is organized as follows. Section~\ref{sec:notations} introduces all the notations and conventions used in the manuscript. 
Section~\ref{sec:theory} provides the dispersion spectrum in the boosted frame with a general hydrodynamic setup. In Section~\ref{sec:MIS}, the boosted dispersion modes are provided for the relativistic MIS theory. Section~\ref{sec:results} includes the results of linearized stability analysis and discusses how it constrains the causality conditions of the theory. In Section~\ref{sec:gen}, we discuss the generality and broader applicability of this method of predicting causal parameter space from linear stability, and we argue that it can be applied to a more general class of low-energy effective field theories. Finally, in Section~\ref{sec:conclu}, we summarize our work and provide possible applications of the current analysis.
%************************
%************************
\section{Notation and convention}
\label{sec:notations}
%************************
%************************
We define $\tilde{\omega}$ and $\tilde{k}$ as the frequency and wavenumber in the local rest frame (LRF). Similarly, $\omega$ and $k$ are the corresponding quantities in a boosted frame, which moves with a uniform background velocity $v$ relative to the LRF. The quantities $u^{\mu}$, $\varepsilon$, and $P$ indicate the hydrodynamic four-velocity, energy density, and pressure of the system. The shear-stress tensor is denoted by $\pi^{\mu\nu}$, with $\eta$ as the shear viscous coefficient and $\tau_{\pi}$ as the relaxation time of the corresponding flow. We have used the abbreviation $\lambda=\eta/\left(\varepsilon_0+P_0\right)$ in our work.
The subscript $0$ denotes the global equilibrium value $\psi_0$ of any field variable $\psi(x)$ where $x^{\mu}$ indicates the space-time co-ordinate.

Throughout this work, we adopt natural units $\hbar = c = k_B = 1$ and use a mostly positive metric signature $g^{\mu\nu} = \{-1,1,1,1\}$.
%************************
%************************
\section{General dispersion structure in a boosted frame}
\label{sec:theory}
%************************
%************************
We begin with a generic expression of the dispersion relation, $\tilde{\omega}=\tilde{\omega}(\tilde{k})$, in the LRF of the fluid. Around the fluid momenta $\tilde k \to 0$, the mode frequency $\tilde \omega$ can be expressed in an infinite series in $\tilde k$ as follows,
\begin{equation}\label{disp-LRF}
    \tilde \omega = \sum_{n=0}^\infty a_n \, \tilde{k}^n~.
\end{equation}
The coefficients $a_n$ can be related to the derivative coefficients of the Taylor expansion of the series given in Eq.~\eqref{disp-LRF}, as $a_n=\frac{1}{n!}\left[\frac{d^n\tilde{\omega}(\tilde{k})}{d\tilde{k}^n}\right]_{\tilde{k}=0}$. These typically consist of constant transport parameters originating from the underlying microscopic theory. If~\eqref{disp-LRF} represents a hydrodynamic mode, we have $a_0=0$~\cite{Bhattacharyya:2025hjs}. Here, without detailed derivation, we are quoting the final results of our previous work~\cite{Bhattacharyya:2025hjs}. We perform the following Lorentz transformation from the LRF ($\tilde{\omega},\tilde{k}$) to a boosted ($\omega,k$) frame with respect to the former,
\begin{align}\label{Lztrans}
    \omega= \gamma\left(\tilde \omega - v \tilde k\right)~,  \quad k =\gamma\left( \tilde k - v \,\tilde \omega \right)~.
\end{align}
In~\cite{Bhattacharyya:2025hjs}, we have shown that for a causal theory, the hydrodynamic and non-hydrodynamic modes of the LRF individually correspond to their respective counterparts in the boosted frame. It means that if we start with a causal theory, no additional modes appear across the boost, or no non-hydrodynamic boosted mode spuriously originates from a LRF hydro mode. For such a well-behaved theory, where the modes have a one-to-one correspondence across Lorentz transformations, we find the following set of boosted dispersion relations solely from the knowledge of the LRF coefficients $a_n$~\cite{Bhattacharyya:2025hjs}. The expression for a boosted hydrodynamic mode is given by,
\begin{equation}\label{boost-hydro}
    \frac{\omega_{h}}{\gamma} = \sum_{n=1}^\infty \tilde a_n \left(\frac{ k}{\gamma}\right)^n~.
\end{equation}
where, the hydro coefficients $\tilde{a}_n$ can be estimated in terms of coefficients $b_n$ by counting the power of $x$ from the
following equation,
\begin{align}\label{tildean}
\sum_{n=0}^\infty \tilde a_{  n+1}x^n = &- v \sum_{n=0}^\infty b_{  n+1} x^n \nonumber\\
&+ \sum_{m=0}^\infty a_{  m+1}x^m\left\{\sum_{l=0}^\infty b_{l+1}x^l \right\}^{m+1}~,       
\end{align}
with $b_n$'s recursively estimated from,
\begin{align}\label{bn}
     \sum_{n=0}^\infty b_{n+1}x^n-v\sum_{m=0}^\infty a_{m+1}x^m\left\{\sum_{l=0}^\infty b_{l+1} x^l \right\}^{m+1} = 1.
\end{align}
The non-hydrodynamic mode in the Lorentz boosted frame is given by,
\begin{align}\label{boost-nonhydro}
    &\gamma\left({\omega_{nh}}+v{k}\right)=\sum_{n=0}^{\infty}\alpha_n^* \left\{\frac{{k}}{\gamma}\right\}^n ~,
\end{align}
where the non-hydro coefficients $\alpha_n^*$ are expressed as \footnote{The notation $a_n^*$ has been used to maintain the parallels with~\cite{Bhattacharyya:2025hjs}, and must not be confused with the complex conjugate of $a_n$.},
\begin{align}\label{LRF-nh}
&\alpha_n^*=\gamma^2 \sum_{m=n}^{\infty}\alpha_m~{}^{m}{\text{C}}_{n}~(  va_0)^{m-n}~,\\
&\text{with,}~~~\alpha_0=\frac{a_0}{\gamma^2}~;~~~~\alpha_1=\tilde{\tilde{a}}_1=\tilde{a}_1+v~;\nonumber\\
&\text{and,}~~~~\alpha_n=\tilde{a}_n~~{\text{for}}~n\geq 2~.
\end{align}
The first few values of the boosted dispersion coefficients in terms of the LRF coefficients $a_n$ are listed as, 
\begin{align}
 \alpha_1=&\tilde{a}_1+v=\frac{1}{\gamma^2}\frac{a_1}{\left(1-a_1 v\right)}~,\\
 \alpha_2=&\tilde{a}_2=\frac{1}{\gamma^2}\frac{a_2}{\left(1-a_1 v\right)^3}~,\\
 \alpha_3=&\tilde{a}_3=\frac{1}{\gamma^2}\frac{a_3 \left(1-a_1 v\right)+2 a_2^2 v}{\left(1-a_1 v\right)^5}~,\\
 \alpha_4=&\tilde{a}_4=\frac{1}{\gamma^2}\left[\frac{a_4}{\left(1-a_1 v\right)^5}\right.\nonumber\\
 &\qquad \qquad + \left.\frac{5 a_2 v \left\{a_2^2 v+a_3 \left(1-a_1 v\right)\right\}}{\left(1-a_1 v\right)^7}\right].
\end{align}
The primary observations from the boosted modes given in~\eqref{boost-hydro} and~\eqref{boost-nonhydro} are the following:
\begin{enumerate}
    \item In the boosted frame, the mode frequency $\omega$ is obtained not merely in a power series of $k$, but rather in a series of $(k/\gamma)$, i.e., expanded in an infinite series of the wavenumber suppressed by the Lorentz factor $\gamma$.
    \item For the non-hydrodynamic modes in~\eqref{boost-nonhydro}, each coefficient $\alpha_n^*$ is itself an infinite series in $(va_0)$. The coefficients of this nested series include the LRF coefficients $a_n$ ranging from $n=0$ to $\infty$, i.e., the knowledge of the LRF modes to their entirety.
\end{enumerate}
%************************
%************************
\section{Boosted dispersion modes of the MIS theory}
\label{sec:MIS}
%************************
%************************
The dispersion spectra of any given theory in the LRF can be obtained by linearizing the governing equations of motion resulting from the conservation laws of the fluid fields around a static background. Here, we consider the conformal MIS theory without any conserved charges. The energy-momentum tensor $T^{\mu\nu}$ for the same is given as,
\begin{align}
T^{\mu\nu}=\varepsilon\left(u^{\mu}u^{\nu}+\frac{1}{3}\Delta^{\mu\nu}\right)+\pi^{\mu\nu},
\end{align}
where the shear stress-tensor $\pi^{\mu\nu}$ satisfies the following equation of motion up to the linear order in gradient corrections,  
\begin{equation}
\left(1+\tau_{\pi}D\right)\pi^{\mu\nu}=-2\eta\sigma^{\mu\nu}~.
\end{equation}
The conservation law $\partial_{\mu}T^{\mu\nu}=0$ gives the necessary equations of motion, which upon linearizing around a static global equilibrium as $\psi(x)=\psi_0+\delta\psi(x)$ and further performing a Fourier analysis in the form
$\psi(x) \sim \int d^4 k ~ e^{-ik_{\mu}x^{\mu}} \tilde{\psi}(k)$, with the wave-vector $k^{\mu}=(\omega,\vec{k})$, will
give the required dispersion polynomials in LRF.
%************************
%************************
\subsection{MIS shear modes}
%************************
%************************
The LRF dispersion polynomial for the shear channel is as follows,
\begin{equation}\label{MIS-shear-poly}
    \tau_{\pi}\tilde{\omega}^2 + i\tilde{\omega} -\lambda\tilde{k}^2 = 0~,
\end{equation}
with $\lambda=\eta/(\varepsilon_0+P_0)$.
Equation~\eqref{MIS-shear-poly} being a quadratic equation, has the two following solutions,
\begin{equation}
\label{quad}
    \tilde{\omega} = \frac{1}{2}\left[-\frac{i}{\tau_\Pi}\pm \sqrt{ -\frac{1}{\tau_\Pi^2} +\frac{4\lambda}{\tau_\Pi}\tilde{k}^2 }\right]\,.
\end{equation}
In order to have an analytic solution for the the dispersion frequency $\tilde{\omega}$ over the $\tilde{k}$ plane, it is instructive to expand $\tilde{\omega}$ in a power series for small $\tilde{k}$. The expansion is supposed to be an infinite series including all derivative orders in $\tilde{k}$ (representing Eq.~\eqref{disp1}), but for practical purposes, we have to truncate it to a certain order in $\tilde{k}$. So, although in principle all the $a_n$ coefficients of Eq.~\eqref{disp1} can be derived, we are presenting the first few of them in the current analysis. Therefore, up to the first few terms, the solutions for the non-hydrodynamic and hydrodynamic modes, respectively, are given by the following series expansion around $\tilde{k}=0$,
\begin{equation}\label{MIS-shear-LRF-nh}
    \tilde{\omega} = -\frac{i}{\tau_\Pi } + i \lambda~\tilde{k}^2 + i \lambda ^2\tau_\Pi~\tilde{k}^4 + 2 i \lambda ^3  \tau_\Pi ^2~ \tilde{k}^6 + {\cal{O}}\left(\tilde{k}^7\right)~,
\end{equation}
and,
\begin{equation}\label{MIS-shear-LRF-hydro}
    \tilde{\omega} = -i \lambda~\tilde{k}^2 - i \lambda ^2  \tau_\Pi~\tilde{k}^4  - 2 i \lambda ^3  \tau_\Pi ^2~\tilde{k}^6 + {\cal{O}}\left(\tilde{k}^7\right)~.
\end{equation}
The conventional method for estimating the modes in a Lorentz-boosted frame is to boost the entire polynomial~\eqref{MIS-shear-poly} and then solve it. Here, bypassing this traditional technique, we derive the boosted modes using the method described in the previous section purely using the expansion coefficients of~\eqref{MIS-shear-LRF-nh} and~\eqref{MIS-shear-LRF-hydro}. Following that, the boosted non-hydrodynamic shear mode can be written as,
% \textcolor{blue}{Should mention here that non-hydro modes can be expressed only as an infinite series expansion, a summation which we haven't been able to perform has we haven't been able to derive all the infinite coefficients, but which can, in principle, be performed with this knowledge. So, we have actually derived the non-hydro modes following the usual method, but the hydro-modes can be exactly derived from our method.} 
% {\color{magenta}{I don't think that much details is really needed here. The results with such details are already documented in JHEP. One can check if wishes. Stating that here will reduce readability and create further confusion.}}
\begin{align}\label{MIS-shear-boost-nh}
    \gamma\left(\omega+vk\right)&=\frac{i}{\left(v^2\lambda-\tau_{\pi}\right)}\left\{\frac{k}{\gamma}\right\}^0
    -\frac{2v\lambda}{\left(v^2\lambda-\tau_{\pi}\right)}\left\{\frac{k}{\gamma}\right\}^1\nn
    &+i\lambda\left\{\frac{k}{\gamma}\right\}^2+2v\lambda^2\left\{\frac{k}{\gamma}\right\}^3\nn
    &-i\lambda^2\left(5v^2\lambda-\tau_{\pi}\right)\left\{\frac{k}{\gamma}\right\}^4+~{\cal{O}}\left\{\frac{k}{\gamma}\right\}^5~.
\end{align}
Similarly, we can evaluate the hydrodynamic shear mode in the boosted frame as,
\begin{align}\label{MIS-shear-boost-hydro}
  \gamma(\omega+vk)&=-i\lambda\left\{\frac{k}{\gamma}\right\}^2-2v\lambda^2\left\{\frac{k}{\gamma}\right\}^3\nn
  &+i\lambda^2\left(5v^2\lambda-\tau_{\pi}\right)\left\{\frac{k}{\gamma}\right\}^4+{\cal{O}}\left\{\frac{k}{\gamma}\right\}^5~.
\end{align}
One can find that the dispersion spectra obtained in~\eqref{MIS-shear-boost-nh} and~\eqref{MIS-shear-boost-hydro} are expressed in power series of $k/\gamma$. The factor $\left(v^2\lambda-\tau_{\pi}\right)$ in the denominator of the leading terms of boosted non-hydro mode~\eqref{MIS-shear-boost-nh} 
comes from the infinite sum over the LRF coefficients given in~\eqref{LRF-nh}.

Here, in~\eqref{MIS-shear-boost-nh} and~\eqref{MIS-shear-boost-hydro}, we have expressed only the first few terms of these expansions for compactness. In Appendix \ref{app:allmodesk6}, we have listed all of these expressions up to $\mathcal{O}(k/\gamma)^6$.
%************************
%************************
\subsection{MIS sound modes}
%************************
%************************
The LRF dispersion polynomial for the MIS sound channel is given as the following,
\begin{equation}\label{MIS-sound-poly}
    \tau_{\pi}\omega^3 + i\omega^2-\left(\frac{4}{3}\lambda+\frac{\tau_{\pi}}{3}\right)\omega k^2 - \frac{i}{3}k^2 = 0~.
\end{equation}
The solution of dispersion polynomial~\eqref{MIS-sound-poly} in the LRF gives the required dispersion series for the MIS sound channel. The first few terms of the non-hydrodynamic mode are given by,
\begin{align}\label{MIS-sound-LRF-nh}
    \tilde{\omega} =& -\frac{i}{\tau_\pi } + i\frac{4}{3}\lambda\tilde{k}^2-i\frac{4}{9}\lambda\tau_\pi(\tau_\pi-4 \lambda )~\tilde{k}^4\nn
    &+i\frac{4}{27}\lambda\tau_\pi ^2\left(32 \lambda ^2-16\lambda\tau_\pi+\tau_\pi^2\right)~ \tilde{k}^6 + {\cal{O}}\left(\tilde{k}^7\right).
\end{align}
Similarly, the hydro modes of the sound channel are expressed as,
\begin{align}\label{MIS-sound-LRF-hydro}
    \tilde{\omega}_{\pm}&=\pm\frac{1}{\sqrt{3}}~\tilde{k}-i\frac{2}{3}\lambda\tilde{k}^2\pm\frac{2\lambda(\tau_\pi-\lambda )}{3 \sqrt{3}}~\tilde{k}^3\nn
    &-i\frac{2}{9}\lambda\tau_\pi (4\lambda-\tau_\pi)~\tilde{k}^4\nn
    &\pm\frac{2\lambda\left(\lambda^3+6\lambda^2\tau_\pi-9\lambda\tau_\pi^2+\tau_\pi^3\right)}{9 \sqrt{3}}~\tilde{k}^5\nn
    &-i\frac{2}{27}\lambda\tau_\pi^2\left(32\lambda^2-16\lambda\tau_\pi+\tau_\pi^2\right)~\tilde{k}^6+{\cal{O}}\left(\tilde{k}^7\right)~.
\end{align}
Following the technique of extracting dispersion spectra at the boosted frame described in the previous section, we find one non-hydro boosted sound mode corresponding to~\eqref{MIS-sound-LRF-nh} as follows,
\begin{align}\label{MIS-sound-boost-nh}
   \gamma\left(\omega+vk\right)&=-\frac{i}{\tau_{\pi}}\frac{\left(1-\frac{v^2}{3}\right)}{\left[1-\frac{v^2}{3}\left(1+\frac{4\lambda}{\tau_{\pi}}\right)\right]}\left\{\frac{k}{\gamma}\right\}^0\nn
   &+\frac{8\lambda}{3\tau_{\pi}}\frac{v}{\left(1-\frac{v^2}{3}\right)\left[1-\frac{v^2}{3}\left(1+\frac{4\lambda}{\tau_{\pi}}\right)\right]}\left\{\frac{k}{\gamma}\right\}^1\nn
   &+i\frac{4\lambda}{3} \frac{(1+v^2)}{\left(1-\frac{v^2}{3}\right)^3}\left\{\frac{k}{\gamma}\right\}^2+{\cal{O}}\left\{\frac{k}{\gamma}\right\}^3~.
\end{align}
The boosted hydro sound modes corresponding to the LRF mode~\eqref{MIS-sound-LRF-hydro} are similarly given by,
\begin{align}\label{MIS-sound-boost-hydro}
&\gamma\left(\omega_{\pm}+vk\right)=\frac{\left(\pm\frac{1}{\sqrt{3}}\right)}{\left(1\mp\frac{v}{\sqrt{3}}\right)}\left\{\frac{k}{\gamma}\right\}-i\frac{2\lambda}{3}\frac{1}{\left(1\mp\frac{v}{\sqrt{3}}\right)^3}\left\{\frac{k}{\gamma}\right\}^2\nn
&~+\frac{2\lambda}{3}\frac{\left[\pm\frac{\left(\tau_{\pi}-\lambda\right)}{\sqrt{3}}-v\left(\lambda+\frac{\tau_{\pi}}{3}\right)\right]}{\left(1\mp\frac{v}{\sqrt{3}}\right)^5}\left\{\frac{k}{\gamma}\right\}^3+{\cal{O}}\left\{\frac{k}{\gamma}\right\}^4.
\end{align}
The sound mode dispersion spectra in~\eqref{MIS-sound-boost-nh} and~\eqref{MIS-sound-boost-hydro} exhibit mode expansion in the power series of $(k/\gamma)$. The factor $\left[1-v^2\left(\frac{1}{3}+\frac{4\lambda}{3\tau_{\pi}}\right)\right]$ in the denominator of the first two leading terms in the non-hydro mode~\eqref{MIS-sound-boost-nh} reflects the infinite sum over the LRF coefficients. We list the expressions of \eqref{MIS-sound-boost-nh} and~\eqref{MIS-sound-boost-hydro} up to $\mathcal{O}(k/\gamma)^6$ in Appendix \ref{app:allmodesk6} for completeness.
%************************
%************************
\section{Results of Linearized Stability Analysis}
\label{sec:results}
%************************
%************************
In this section, we discuss the results of linearized stability analysis of the non-hydrodynamic modes derived in the previous section using the frame-invariant stability criterion ${\rm Im} (\omega) \leq |{\rm Im}(k)|$~\cite{Gavassino:2023myj}. Since the expressions for $\omega$ are generally calculated only up to a finite order in $k$, the results of the stability analysis are also likely to have limited accuracy. Nevertheless, we will see that the stability analysis results for a boosted frame with $v\to 1$ are exact despite the limited accuracy of $\omega$, owing to the ``$\gamma$-suppression" behavior of the dispersion relation~\cite{Bhattacharyya:2025hjs}. This phenomenon sharply drops off the next-to-leading order terms, making them quantitatively insignificant beyond the leading term. This is the main motivation of this analysis that finally leads to identifying the causal parameter space. ~\footnote{One could also try to perform a Routh-Hurwitz stability check of the $\omega$'s directly from the dispersion polynomial itself, as we previously did in~\cite{Roy:2023apk}, but due to the generically complex-valued $k$'s, this analysis becomes significantly difficult now. We leave this work for our future endeavors.}

Here, we discuss only the non-hydrodynamic modes because the information encoded in these modes governs the causality properties of the theory, since the very existence of such modes is required by causality~\cite{Hoult:2023clg,Heller:2022ejw}. In this work, we are interested in showing how the question of causality, which is known to be a high-energy property of the theory, can be addressed while staying within the low-energy regime of the theory, using the exclusive nature of the $v \to 1, k \to 0$ stability property due to the $\gamma$-suppression of the boosted modes.
%************************
%************************
\subsection{Results of stability analysis at \texorpdfstring{$k=0$}{}: a review}
%************************
%************************
We begin this section with a review of our previous results of the MIS theory discussed in~\cite{Roy:2023apk}. In order to investigate the stability conditions at arbitrary boost $v$ and $k\to 0$ limit, in Ref.~\cite{Roy:2023apk} the Routh-Hurwitz (R-H) stability test was used for higher-order dispersion polynomials. Here, to maintain uniformity with the current work, we use the stability condition ${\rm Im}(\omega) \leq |{\rm Im}(k)| $ at the spatially homogeneous limit $k \to 0$, which translates to the inequality ${\rm Im}(\omega)|_{k\to0} \leq 0$, and essentially gives the same result as R-H test. 

\begin{figure*}[ht!]
    \centering
    \begin{subfigure}{0.2\textwidth}
        \centering
        \includegraphics[width=\textwidth]{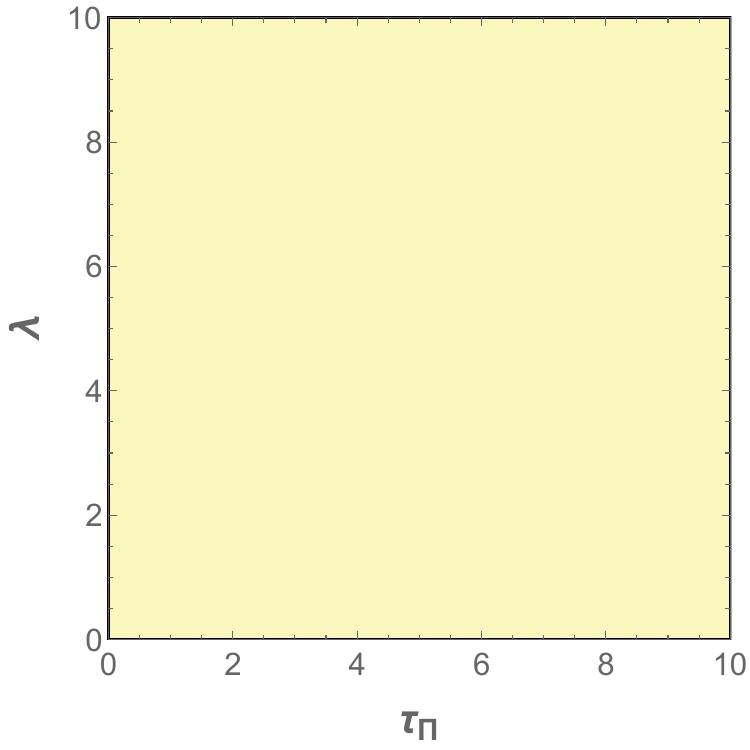}
        \caption{$v=0$}
        \label{fig:Shear:k0v0}
    \end{subfigure}
    \hfill % Adds horizontal space
    \begin{subfigure}{0.2\textwidth}
        \centering
        \includegraphics[width=\textwidth]{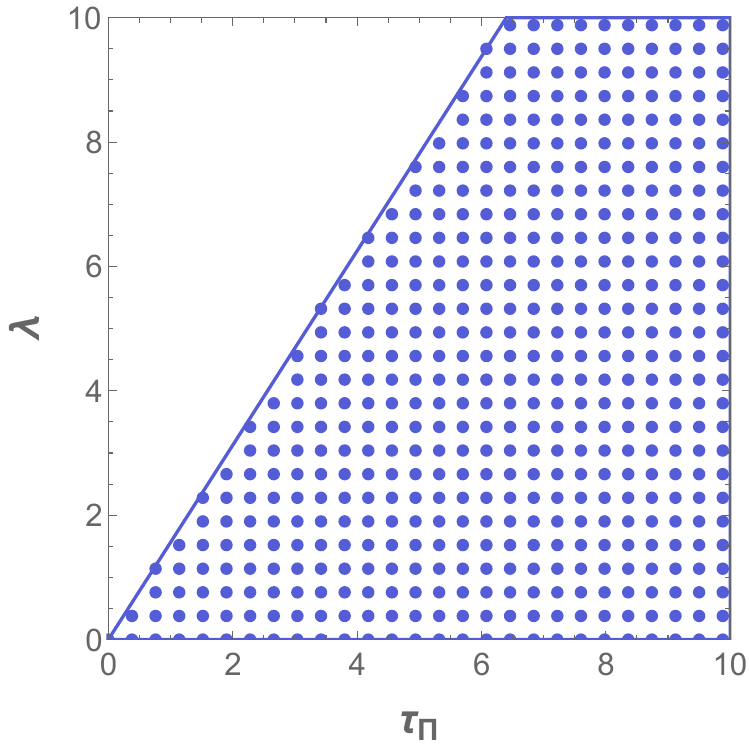}
        \caption{$v=0.8$}
        \label{fig:Shear:k0v08}
    \end{subfigure}
    \hfill
    \begin{subfigure}{0.2\textwidth}
        \centering
        \includegraphics[width=\textwidth]{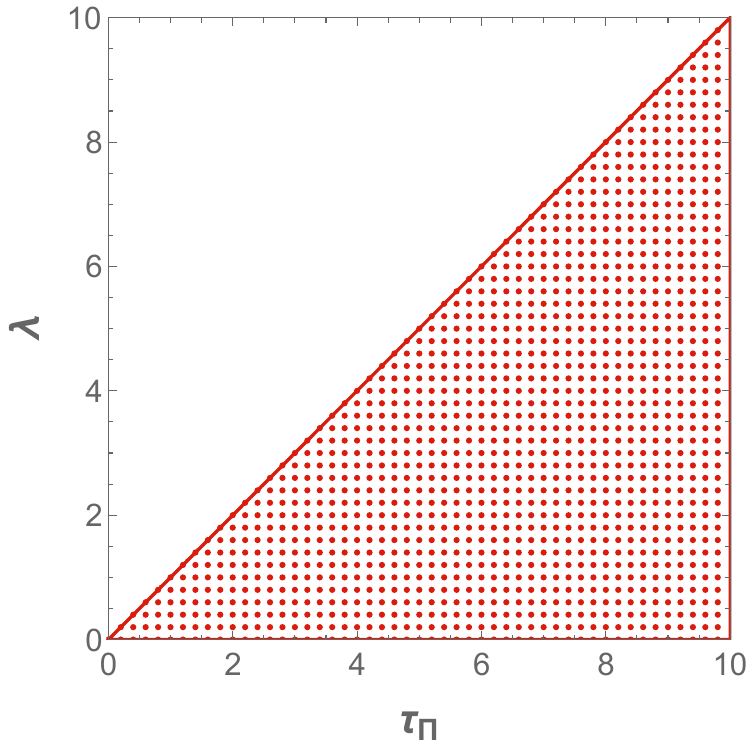}
        \caption{$v=0.999$}
        \label{fig:Shear:k0v1}
    \end{subfigure}
    \hfill
    \begin{subfigure}{0.2\textwidth}
        \centering
        \includegraphics[width=\textwidth]{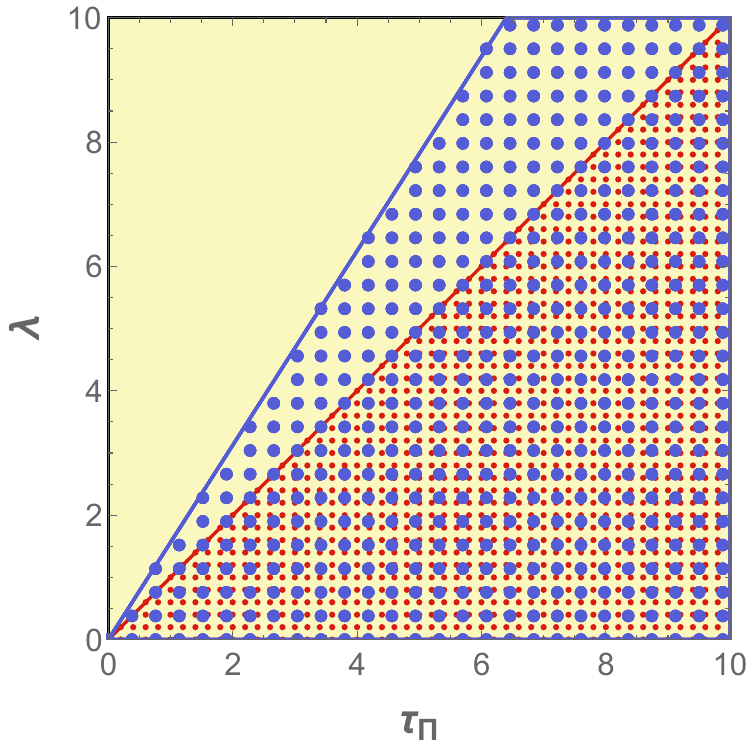}
        \caption{Overlap of \ref{fig:Shear:k0v0}, \ref{fig:Shear:k0v08}, \ref{fig:Shear:k0v1}}
        \label{fig:Sheark0voverlap}
    \end{subfigure}
    \caption{Parameter space satisfying ${\rm Im}(\omega)\leq |{\rm Im}(k)|$ for different boosts at $O(k^0)$ in the Shear channel.}
    \label{fig:Shear:k0}
\end{figure*}

\begin{figure*}[ht!]
    \centering
    \begin{subfigure}{0.2\textwidth}
        \centering
        \includegraphics[width=\textwidth]{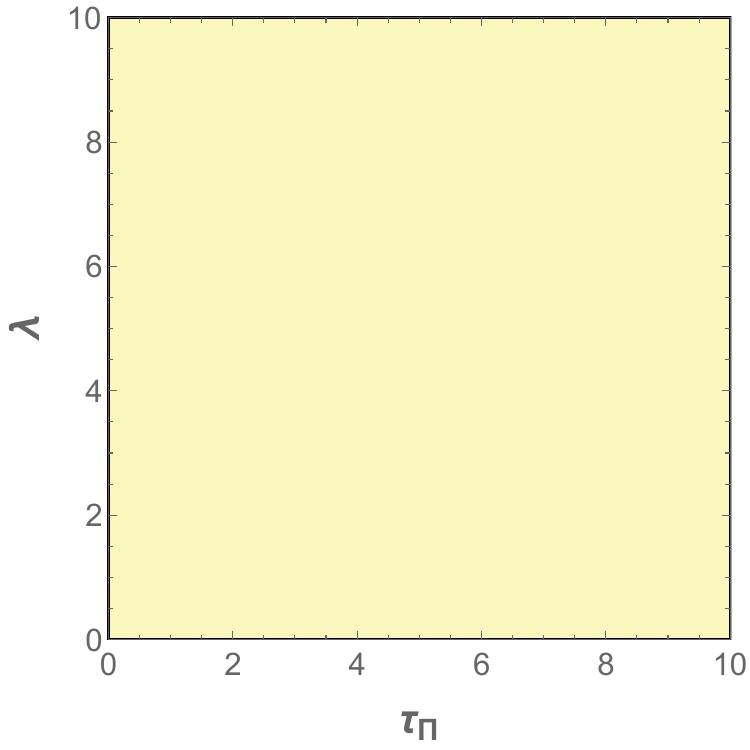}
        \caption{$v=0$}
        \label{fig:Sound:k0v0}
    \end{subfigure}
    \hfill % Adds horizontal space
    \begin{subfigure}{0.2\textwidth}
        \centering
        \includegraphics[width=\textwidth]{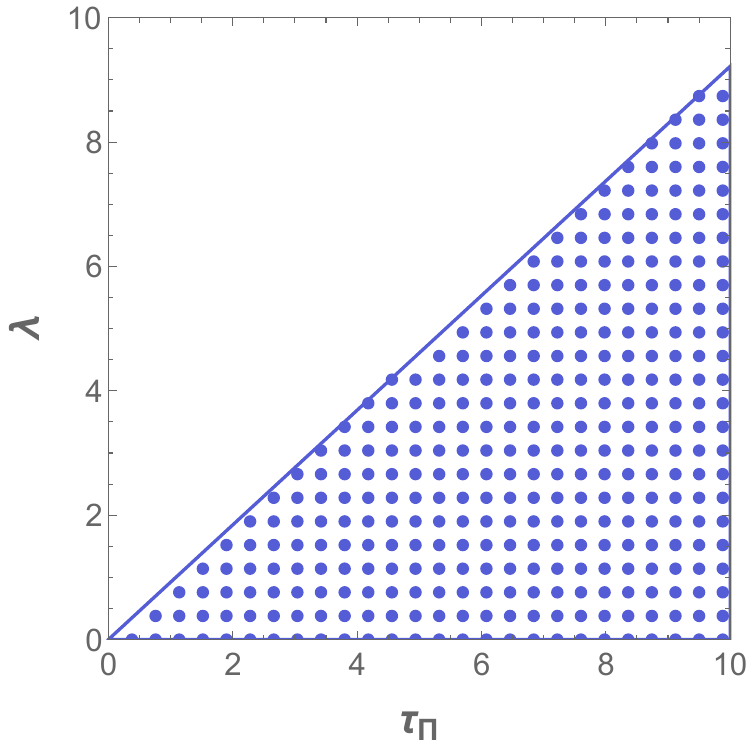}
        \caption{$v=0.8$}
        \label{fig:Sound:k0v08}
    \end{subfigure}
    \hfill
    \begin{subfigure}{0.2\textwidth}
        \centering
        \includegraphics[width=\textwidth]{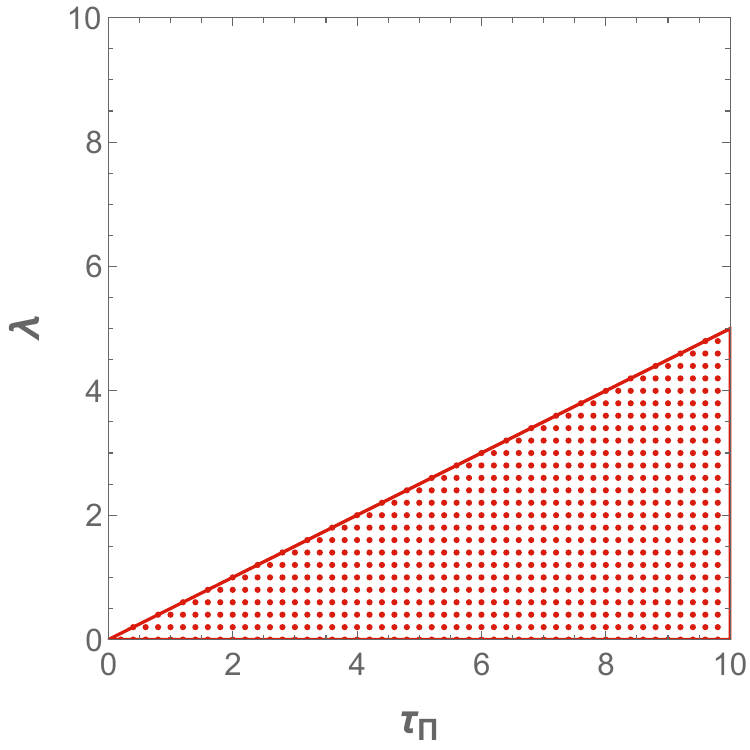}
        \caption{$v=0.999$}
        \label{fig:Sound:k0v1}
    \end{subfigure}
    \hfill
    \begin{subfigure}{0.2\textwidth}
        \centering
        \includegraphics[width=\textwidth]{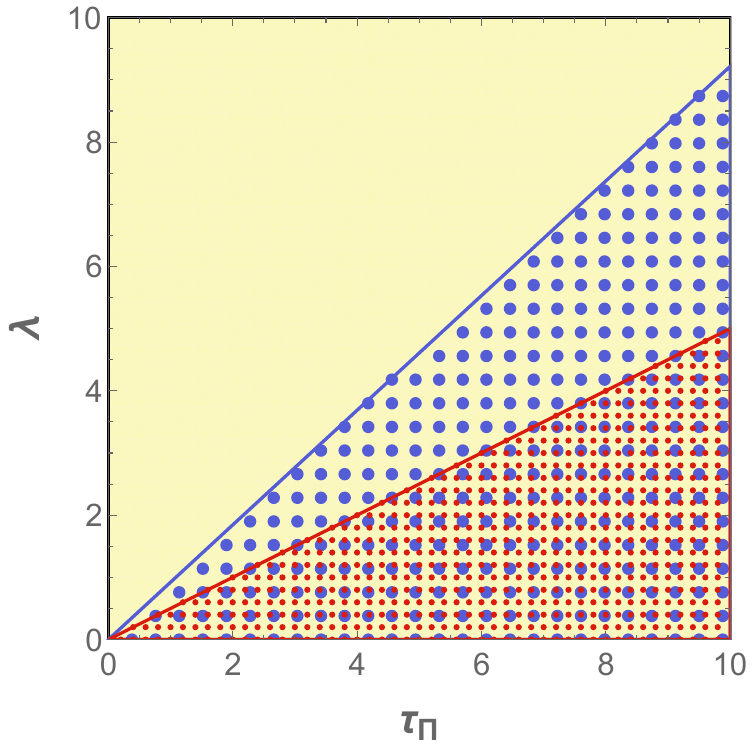}
        \caption{Overlap of \ref{fig:Sound:k0v0}, \ref{fig:Sound:k0v08}, \ref{fig:Sound:k0v1}}
        \label{fig:Soundk0voverlap}
    \end{subfigure}
    \caption{Parameter space satisfying ${\rm Im}(\omega)\leq |{\rm Im}(k)|$ for different boosts at $O(k^0)$ in the Sound channel.}
    \label{fig:Sound:k0}
\end{figure*}

The mentioned stability criteria for both the shear and sound channels of the MIS theory have been plotted in the $\lambda-\tau_\pi$ parameter space for $k \to 0$ and different boost velocities $v$. They are respectively depicted in Figs.~\ref{fig:Shear:k0} and \ref{fig:Sound:k0}, where the shaded regions indicate the domain of stability. By observing how this ``parameter space of stability" (called $PSS$ hereafter) changes with changing $v$, we notice several interesting features.

At $v=0$, i.e., at the LRF of the fluid, the theory is stable over the full available $\lambda-\tau_\pi$ parameter space as long as both $\lambda,\tau_{\pi}$ are positive. Increasing the boost velocity $v$ from $0$ to $0.999$ \footnote{For near-luminal velocity, we restrict our analysis to $v=0.999$, which is sufficiently close to the $v \to 1$ limit that we are interested in. One could always use a higher upper limit of $v$, e.g., $v=0.9999$, but we find that any modifications to the plots under such a change are barely noticeable.}, we find that the $PSS$ decreases monotonically with increasing boost.\footnote{Although the full range of $v$ is $-1 \leq v \leq 1$, the dispersion polynomial at $k \to 0$ is symmetric under $v \to -v$. Therefore, it suffices to perform the analysis in the range $0 \leq v \leq 1$.} This reduction in the $PSS$ with increasing $v$ occurs in such a way that the $PSS$ at a value of $v$ (say $v_1$) is always enclosed within the $PSS$ of any $v$ (say $v_2$) that is lesser than $v_1$, i.e., $$PSS (v_1) \subset PSS (v_2),~ \forall ~v_2 < v_1~.$$
This leads us to conclude that the $PSS$ at the maximum value of $v$ (within the allowed range $0\leq v<1$) will be enclosed within that of all other values of $v$, i.e., $$ PSS (v\to 1) \subset PSS (v), ~ \forall ~ 0 \leq v < 1 ~.$$
Following the argument that a dissipative theory which is stable in all reference frames must be causal~\cite{Gavassino:2023myj}, the $PSS (v\to 1)$ gives us the parameter space where the theory is causal, or the ``parameter space of causality" ($PSC$) \footnote{Since we have restricted ourselves only to linearized stability analysis, any statement related to causality will also be made under this assumption, that it can be derived from a linearized analysis.}.
% \begin{widetext}
% \end{widetext} 

At this juncture, it is worth discussing the idea of necessary and sufficient conditions. Event A is a sufficient condition for Event B if, for every occurrence of Event A, there is an occurrence of Event B. Event A is a necessary condition for the occurrence of Event B if there is no occurrence of Event B for which Event A does not occur.

Now, let us analyze this derived causal parameter space a little further. We find that every point outside the $PSS (v\to 1)$ violates this stability-invariance condition of causality (see Fig.~\ref{fig:Shear:k0} and \ref{fig:Sound:k0}), i.e., there always exists some value of $v$ beyond which the stability criteria are not satisfied at such a point, and the stability-invariance breaks down. This shows that the $PSS(v \to 1)$ gives us the necessary condition for causality. Conversely, every point lying within $PSS(v \to 1)$ satisfies the stability criteria in all reference frames; thus, it is frame-invariantly stable and, hence, causal. Therefore, the $PSS(v \to 1)$ also gives us the sufficient condition of causality as well. Hence, the stability criteria at $v \to 1$ give us the necessary and sufficient constraint for linearized causality at the spatially homogeneous limit. Incidentally, it turns out that the causality conditions derived from the near-luminal, frame-invariant stability criteria coincide with the asymptotic causality condition $\lim _{k \to \infty} | \frac{\partial \omega}{\partial k}| < 1$, as shown in~\cite{Roy:2023apk}.

In Figs.~\ref{fig:Shear:k0} and \ref{fig:Sound:k0}, we have plotted the $PSS$ for three different values of $v$ for the MIS shear and sound channel, respectively. Figures.~\ref{fig:Shear:k0v0} and \ref{fig:Sound:k0v0} show the $PSS(v=0)$, Figs. \ref{fig:Shear:k0v08} and \ref{fig:Sound:k0v08} show the $PSS$ at an intermediate value of the boost at $v=0.8$, and Figs.~\ref{fig:Shear:k0v1} and \ref{fig:Sound:k0v1} show the $PSS(v \to 1)$ (with $v=0.999$), for the MIS shear and sound channel, respectively. The overlaps of these plots for different boost velocities, as displayed in Figs.~\ref{fig:Sheark0voverlap} and \ref{fig:Soundk0voverlap} show that indeed $PSS (v\to 1)$ is enclosed within the $PSS(v=0)$ and $PSS(v=0.8)$ for both channels, supporting the claims we made.

Motivated by the results obtained at $k=0$, the questions we would like to address next are the following: \textit{(i)} What happens at $k \neq 0$? \textit{(ii)} Does this method of extracting causal parameter space from stability invariance still hold at non-zero momenta? \textit{(iii)} Do the stability criteria at $v \to 1$ still lead us to the necessary and sufficient criteria for causality? 

The answers to these questions are the key findings of the current work, which we present in the following discussions.
%************************
%************************
\subsection{Results at \texorpdfstring{$k\neq 0$}{}}
\label{subsec:kneq0}
%************************
%************************
Following the same procedure as before, we now use the non-zero $k$ expanded series of $\omega$ from Eq.~\eqref{MIS-shear-boost-nh} and~\eqref{MIS-sound-boost-nh} up to $\mathcal{O}(k^6)$. Varying the value of $v$ from $0$ to $0.999$, we investigate how the $PSS$ changes for both the shear and sound channels for several non-zero, complex $k$ values.~\footnote{With $k \neq 0$, the dispersion polynomial is no longer symmetric under $v \to -v$.
However, the qualitative results and conclusions, i.e., the non-monotonic change of the $PSS$, and the matching of the $PSS(v\to 1,k\neq0)$ with the $PSS(v\to 1, k =0)$, remain the same.}
% \begin{widetext}
    \begin{figure*}[ht!]
    \centering
    \begin{subfigure}{0.2\textwidth}
        \centering
        \includegraphics[width=\textwidth]{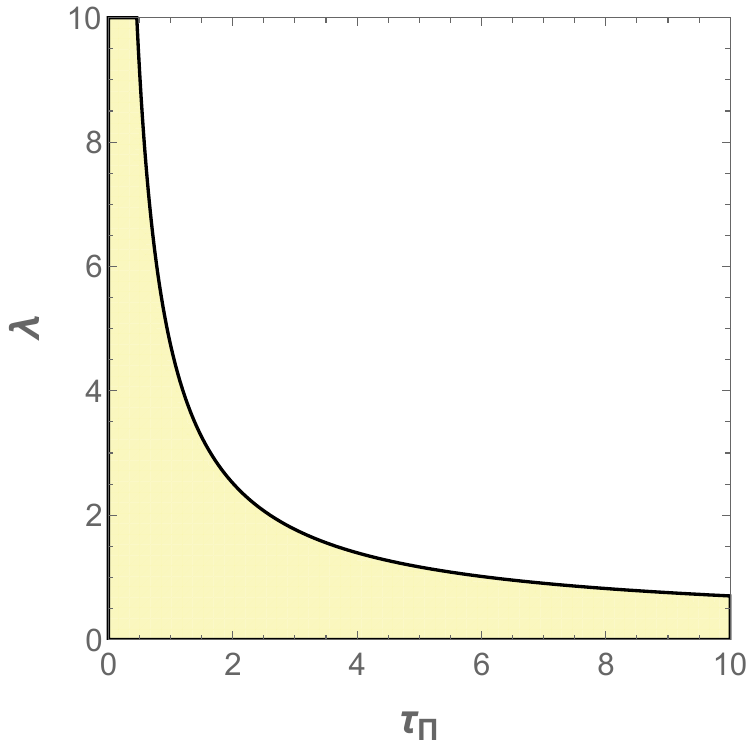}
        \caption{$v=0$}
        \label{fig:Shear:kneq0v0}
    \end{subfigure}
    \hfill % Adds horizontal space
    \begin{subfigure}{0.2\textwidth}
        \centering
        \includegraphics[width=\textwidth]{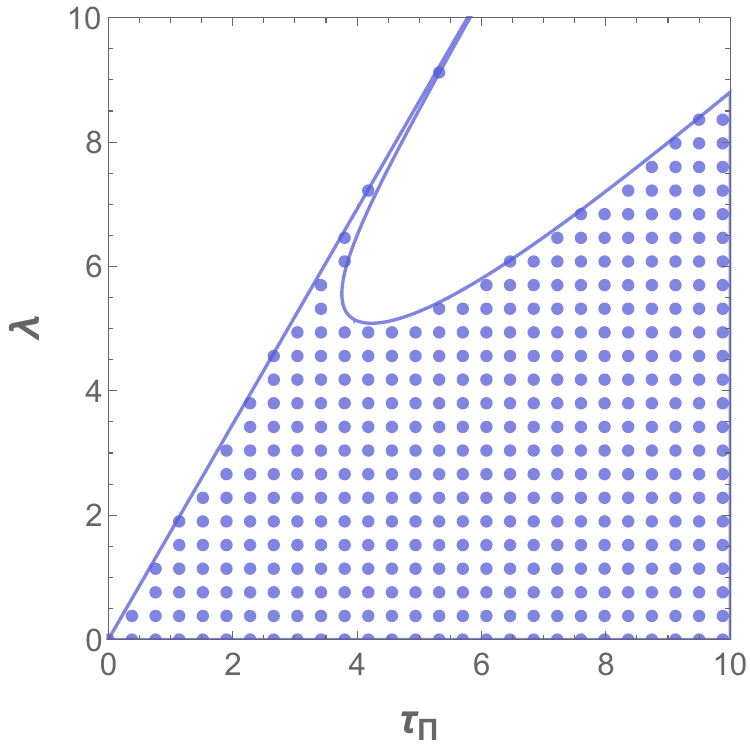}
        \caption{$v=0.76$}
        \label{fig:Shear:kneq0v076}
    \end{subfigure}
    \hfill
    \begin{subfigure}{0.2\textwidth}
        \centering
        \includegraphics[width=\textwidth]{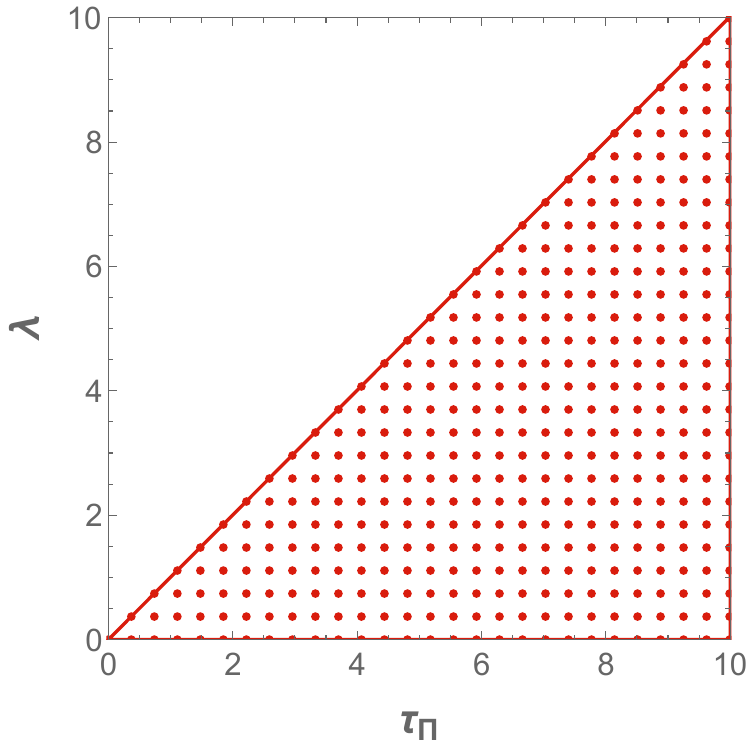}
        \caption{$v=0.999$}
        \label{fig:Shear:kneq0v1}
    \end{subfigure}
    \hfill
    \begin{subfigure}{0.2\textwidth}
        \centering
        \includegraphics[width=\textwidth]{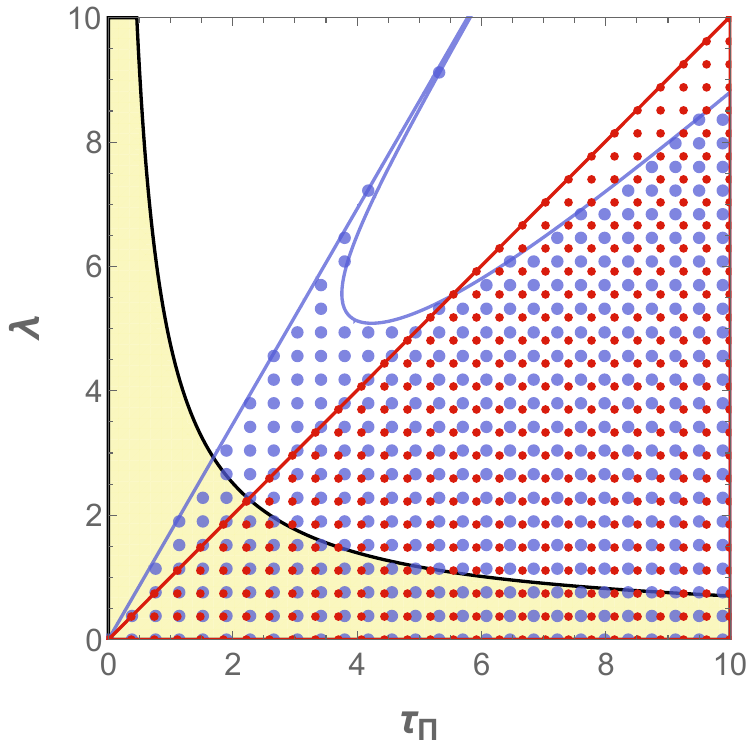}
        \caption{Overlap of \ref{fig:Shear:kneq0v0}, \ref{fig:Shear:kneq0v076}, \ref{fig:Shear:kneq0v1}}
        \label{fig:Shear:kneq0voverlap}
    \end{subfigure}
    \caption{Parameter space satisfying ${\rm Im}(\omega)\leq |{\rm Im}(k)|$ for different boosts at $O\left(k^6\right)$ with $k = 0.394+ 0.1~i$ in the Shear channel.}
    \label{fig:Shear:kneq0}
\end{figure*}
\begin{figure*}[ht!]
    \centering
    \begin{subfigure}{0.2\textwidth}
        \centering
        \includegraphics[width=\textwidth]{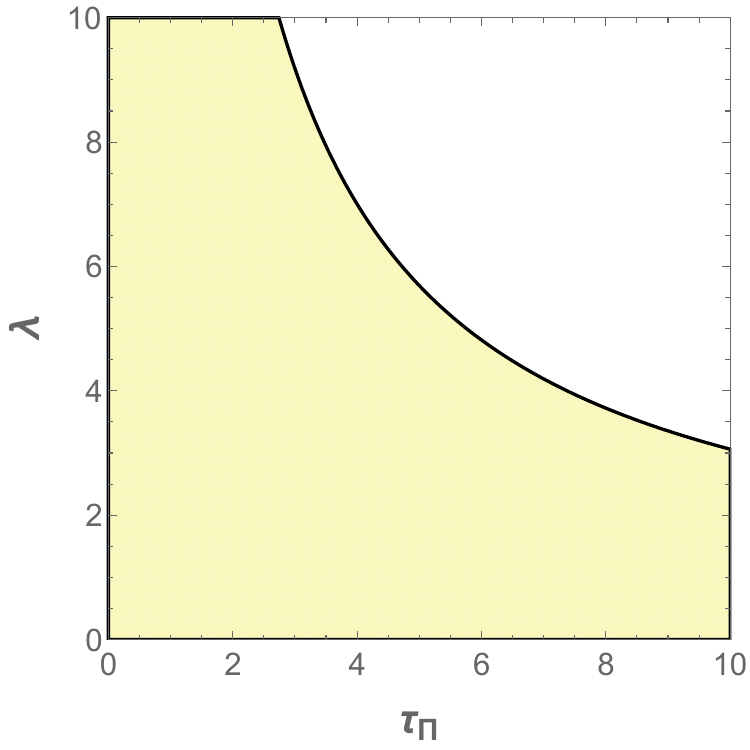}
        \caption{$v=0$}
        \label{fig:Shear:kneq0bv0}
    \end{subfigure}
    \hfill % Adds horizontal space
    \begin{subfigure}{0.2\textwidth}
        \centering
        \includegraphics[width=\textwidth]{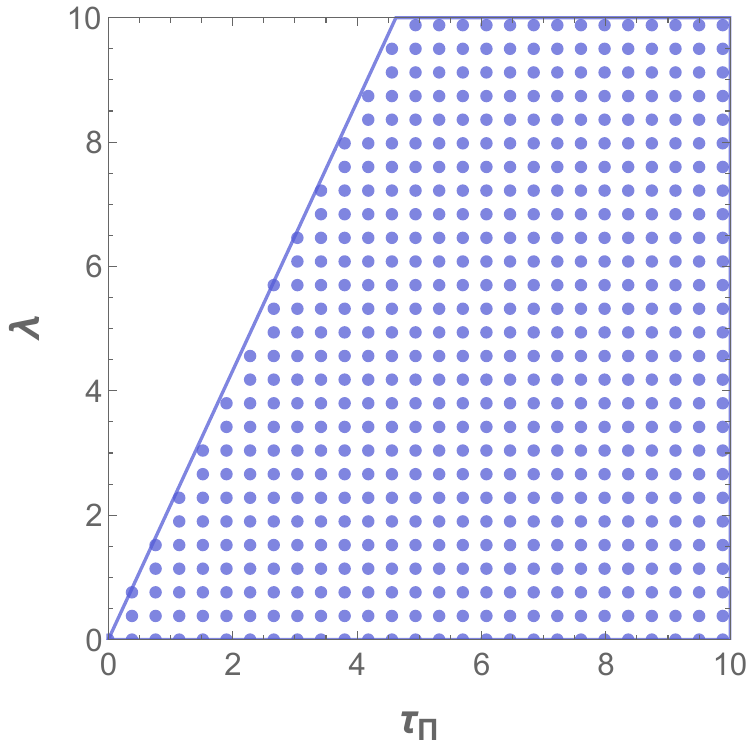}
        \caption{$v=0.68$}
        \label{fig:Shear:kneq0bv068}
    \end{subfigure}
    \hfill
    \begin{subfigure}{0.2\textwidth}
        \centering
        \includegraphics[width=\textwidth]{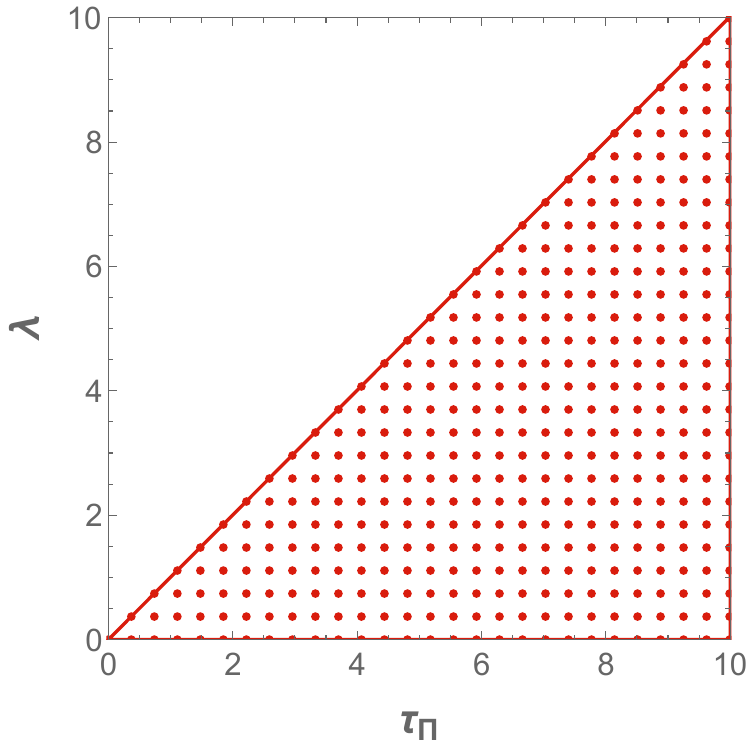}
        \caption{$v=0.999$}
        \label{fig:Shear:kneq0bv1}
    \end{subfigure}
    \hfill
    \begin{subfigure}{0.2\textwidth}
        \centering
        \includegraphics[width=\textwidth]{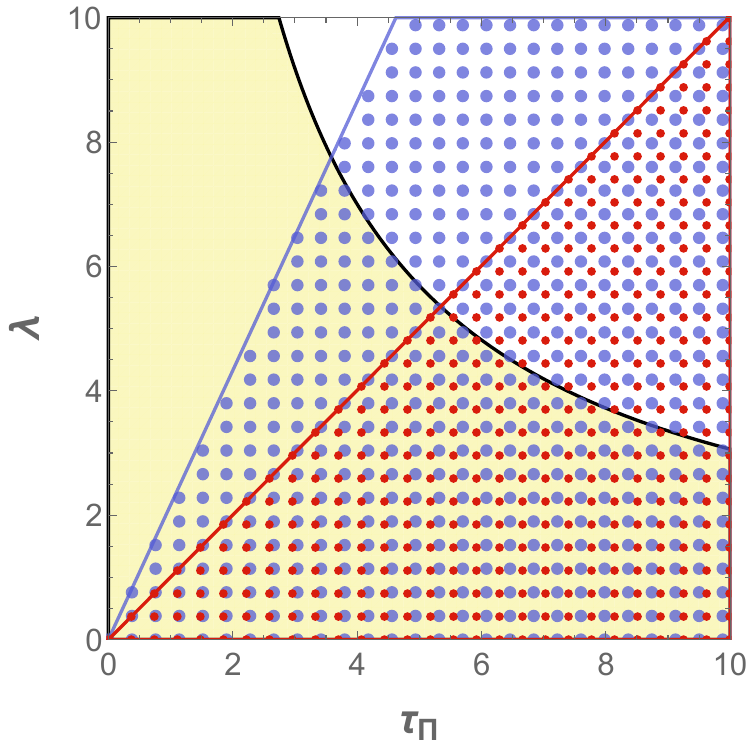}
        \caption{Overlap of \ref{fig:Shear:kneq0bv0}, \ref{fig:Shear:kneq0bv068}, \ref{fig:Shear:kneq0bv1}}
        \label{fig:Shear:kneq0bvoverlap2}
    \end{subfigure}
    \caption{Parameter space satisfying ${\rm Im}(\omega)\leq |{\rm Im}(k)|$ for different boosts at $O\left(k^6\right)$ with $k = 0.15+ 0.03~i$ in the Shear channel.}
    \label{fig:Shear:kneq0b}
\end{figure*}
    \begin{figure*}[ht!]
    \centering
    \begin{subfigure}{0.2\textwidth}
        \centering
        \includegraphics[width=\textwidth]{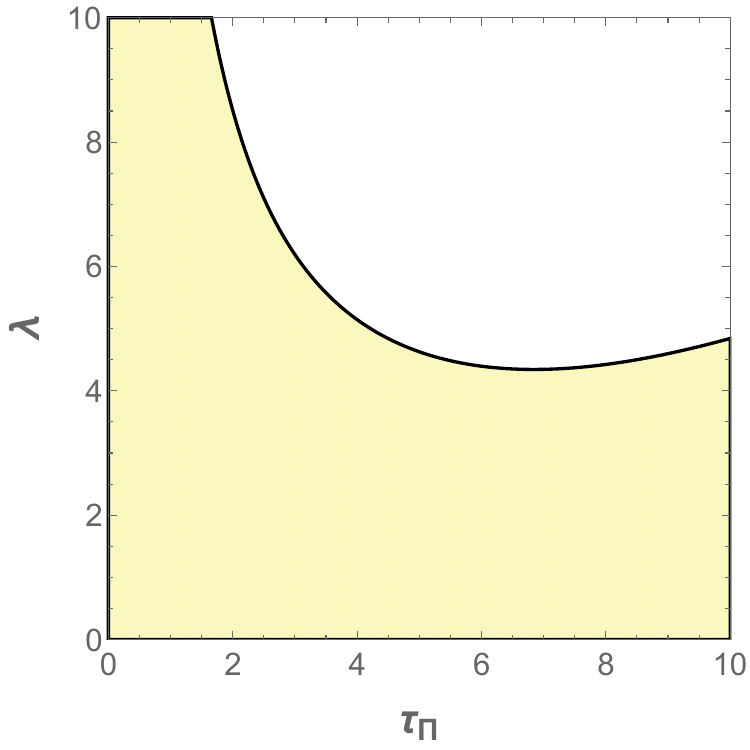}
        \caption{$v=0$}
        \label{fig:Sound:kneq0v0}
    \end{subfigure}
    \hfill % Adds horizontal space
    \begin{subfigure}{0.2\textwidth}
        \centering
        \includegraphics[width=\textwidth]{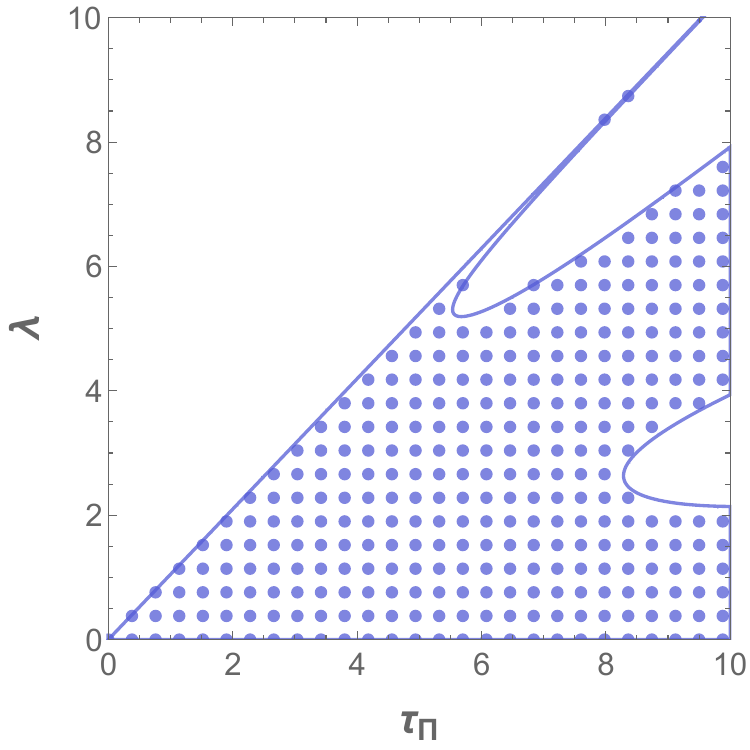}
        \caption{$v=0.76$}
        \label{fig:Sound:kneq0v076}
    \end{subfigure}
    \hfill
    \begin{subfigure}{0.2\textwidth}
        \centering
        \includegraphics[width=\textwidth]{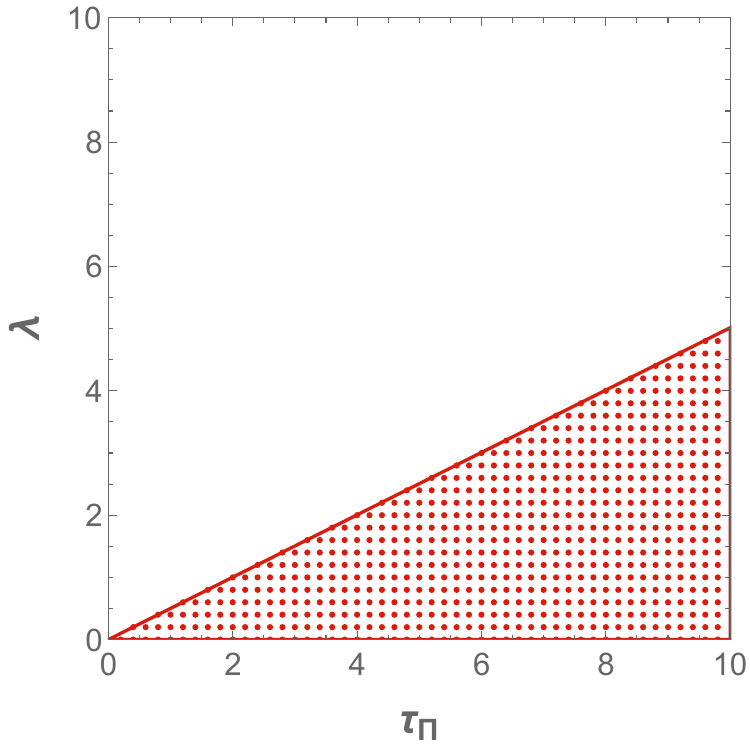}
        \caption{$v=0.999$}
        \label{fig:Sound:kneq0v1}
    \end{subfigure}
    \hfill
    \begin{subfigure}{0.2\textwidth}
        \centering
        \includegraphics[width=\textwidth]{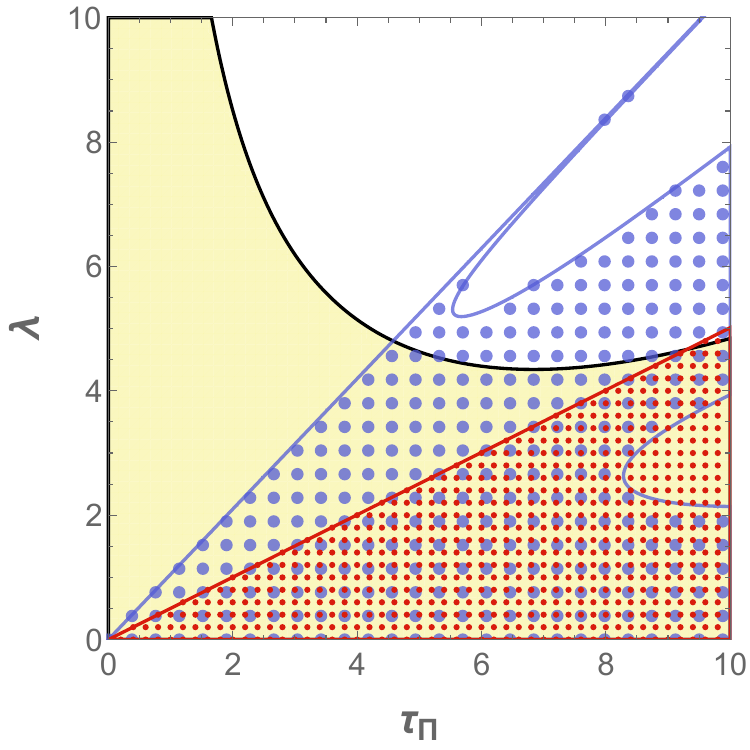}
        \caption{Overlap of \ref{fig:Sound:kneq0v0}, \ref{fig:Sound:kneq0v076}, \ref{fig:Sound:kneq0v1}}
        \label{fig:Sound:kneq0voverlap}
    \end{subfigure}
    \caption{Parameter space satisfying ${\rm Im}(\omega)\leq |{\rm Im}(k)|$ for different boosts at $O\left(k^6\right)$ with $k = (0.1+ 0.2~i)$ in the Sound channel.}
    \label{fig:Sound:kneq0}
\end{figure*}
\begin{figure*}[ht!]
    \centering
    \begin{subfigure}{0.2\textwidth}
        \centering
        \includegraphics[width=\textwidth]{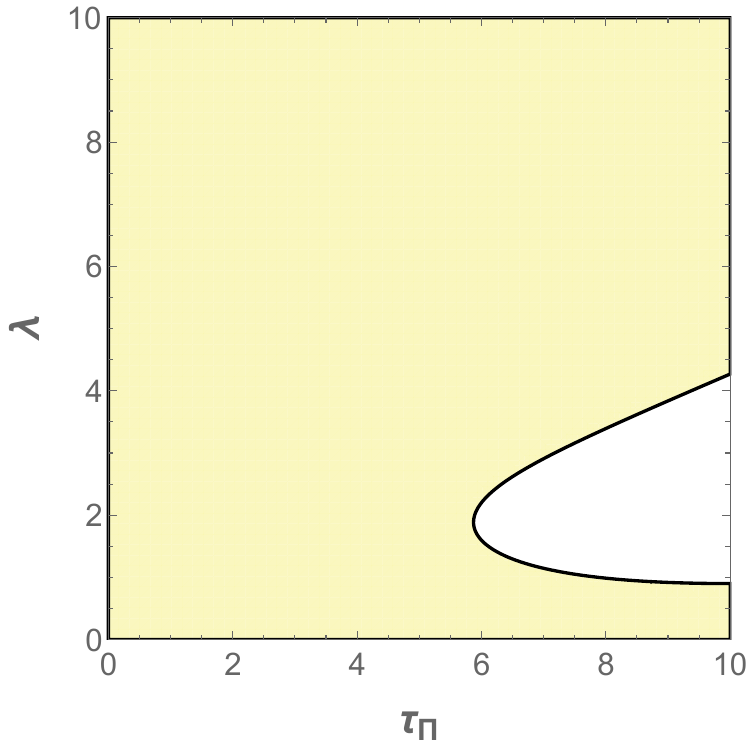}
        \caption{$v=0$}
        \label{fig:Sound:kneq0bv0}
    \end{subfigure}
    \hfill % Adds horizontal space
    \begin{subfigure}{0.2\textwidth}
        \centering
        \includegraphics[width=\textwidth]{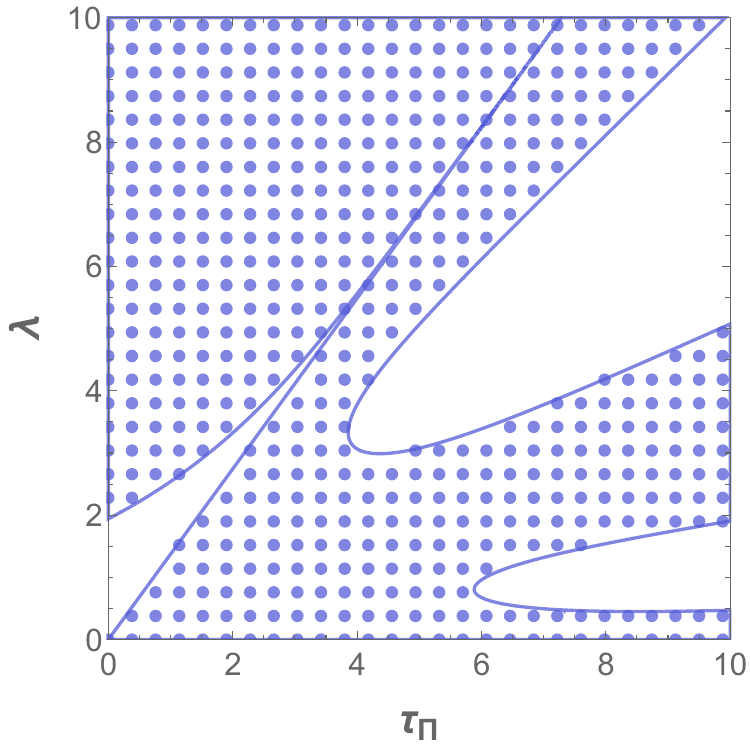}
        \caption{$v=0.68$}
        \label{fig:Sound:kneq0bv068}
    \end{subfigure}
    \hfill
    \begin{subfigure}{0.2\textwidth}
        \centering
        \includegraphics[width=\textwidth]{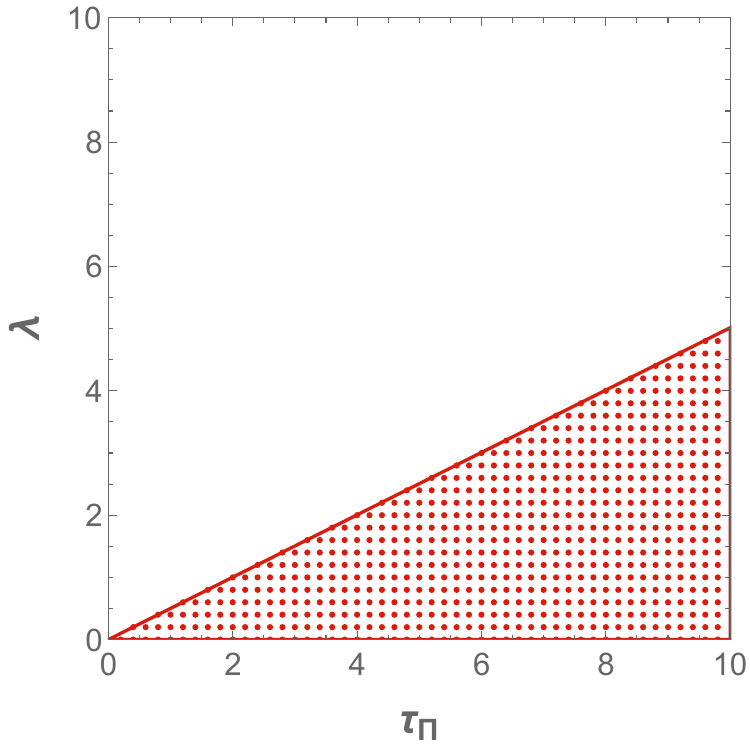}
        \caption{$v=0.999$}
        \label{fig:Sound:kneq0bv1}
    \end{subfigure}
    \hfill
    \begin{subfigure}{0.2\textwidth}
        \centering
        \includegraphics[width=\textwidth]{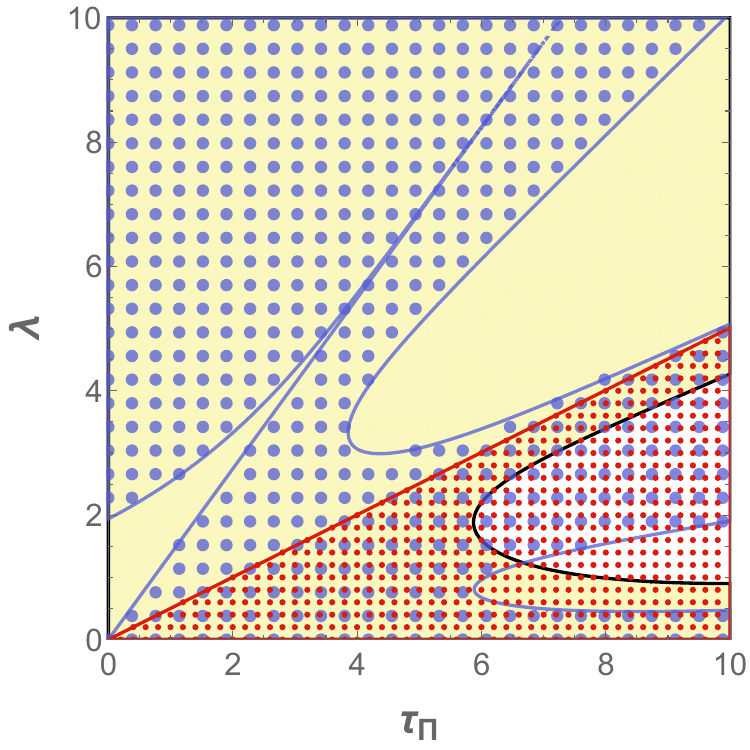}
        \caption{Overlap of \ref{fig:Sound:kneq0bv0}, \ref{fig:Sound:kneq0bv068}, \ref{fig:Sound:kneq0bv1}}
        \label{fig:Sound:kneq0bvoverlap2}
    \end{subfigure}
    \caption{Parameter space satisfying ${\rm Im}(\omega)\leq |{\rm Im}(k)|$ for different boosts at $O\left(k^6\right)$ with $k = (0.27+ 0.12~i)$ in the Sound channel.}
    \label{fig:Sound:kneq0b}
\end{figure*}
%\end{widetext}
%\begin{widetext}

% \end{widetext}

In Figs.~\ref{fig:Shear:kneq0}, \ref{fig:Shear:kneq0b}, \ref{fig:Sound:kneq0} and \ref{fig:Sound:kneq0b}, we have plotted the $PSS$ with three different values of $v$ for each of two different values of $k$ for the MIS shear and sound channels, respectively, along with their overlap regions. 
Compared to the $k=0,~v=0$ case, we now find a substantially reduced $PSS$ on the $\lambda-\tau_\pi$ plane at $k \neq 0,~v=0$. Increasing $v$ from 0, we find that the $PSS$ changes as expected and shifts from the $\lambda$ axis but never leaves the $\tau_{\pi}$ axis. The analytical reason behind this behavior has been derived in Appendix-\ref{appen-2}. However, the monotonic behavior with increasing $v$ as observed in the spatially homogeneous case is lost. The key observation here is that, at $v\rightarrow 1$, it matches with the exact same parameter space as for the $k=0$ case, irrespective of the non-zero $k$ values taken. 

A close comparison with Fig.~\ref{fig:Shear:k0v1} and \ref{fig:Sound:k0v1} shows that at $v\rightarrow 1$ the stability analysis is insensitive to the $k$ values. It is true that the $PSS$ is no longer entirely enclosed within those of all lower values of $v$. From the overlap plots in Fig.~\ref{fig:Shear:kneq0voverlap}, \ref{fig:Shear:kneq0bvoverlap2}, \ref{fig:Sound:kneq0voverlap} and \ref{fig:Sound:kneq0bvoverlap2}, it is clear that some portion of the $PSS (v\to 1)$ is not enclosed within the $PSS$ of other boosts, which indicates that unlike $k=0$ case, now the entire $PSS (v\to 1)$ is not frame-invariantly stable. This somehow restricts us from predicting a sufficient causal parameter space from frame-invariant stability arguments at $k \neq 0$. But the crucial observation is that, for all the non-zero $k$ values shown here, the common overlap zone of all boosts (see Fig.~\ref{fig:Shear:kneq0voverlap} to \ref{fig:Sound:kneq0bvoverlap2}), i.e., the frame-invariantly stable parameter space is necessarily enclosed within the near-luminal $PSS$ of the given momenta, which by virtue of $\gamma$-suppression reduces to $PSS$($k=0,v\rightarrow 1$). 

Consequently, following the chain of arguments above, it yields the parameter space of the theory in which the necessary condition of causality is satisfied. This is the \textit{main result} of the current analysis, that even at non-zero momenta, the near-luminal parameter space at $k=0$ suffices to provide the necessary condition for causality in terms of the transport coefficients of the theory that set the parameter space. 
The same behavior is exhibited for different sets of $k$-values for both the shear and the sound channels in Fig. \ref{fig:Shear:kneq0} to \ref{fig:Sound:kneq0b}.

    \begin{figure*}[ht!]
    \centering
    \begin{subfigure}{0.3\textwidth}
        \centering
        \includegraphics[width=\textwidth]{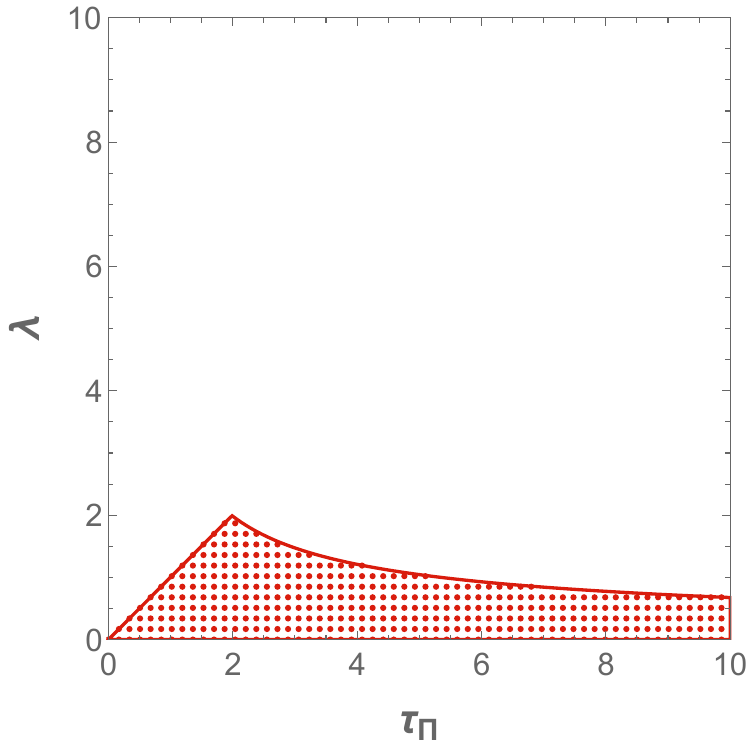}
        \caption{$\bigcap_{v \in (-1,1)} PSS(v)$}
        \label{fig:Shear:suffvall}
    \end{subfigure}
    \hfill
    \begin{subfigure}{0.3\textwidth}
        \centering
        \includegraphics[width=\textwidth]{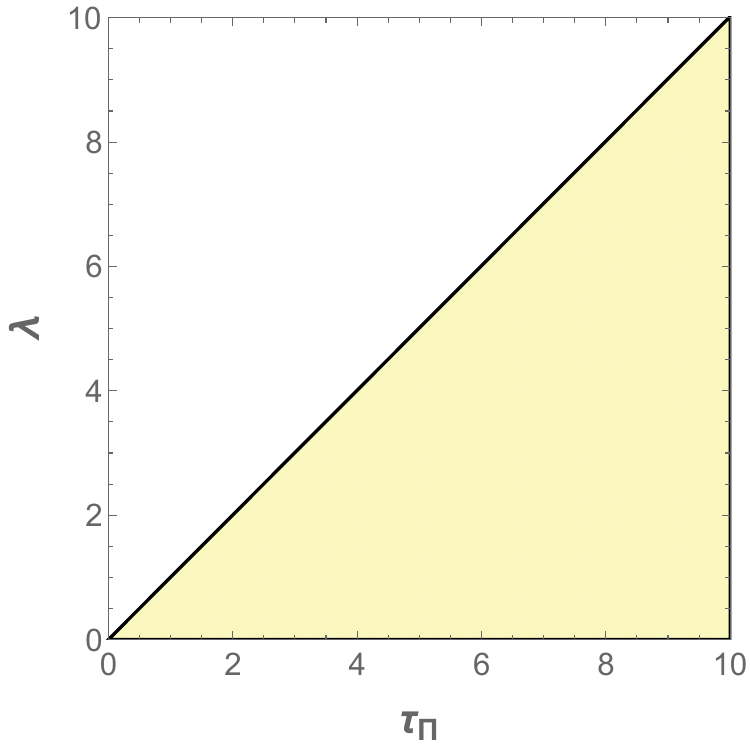}
        \caption{$PSS(v \to 1)$}
        \label{fig:Shear:suffv1}
    \end{subfigure}
    \hfill % Adds horizontal space
    \begin{subfigure}{0.3\textwidth}
        \centering
        \includegraphics[width=\textwidth]{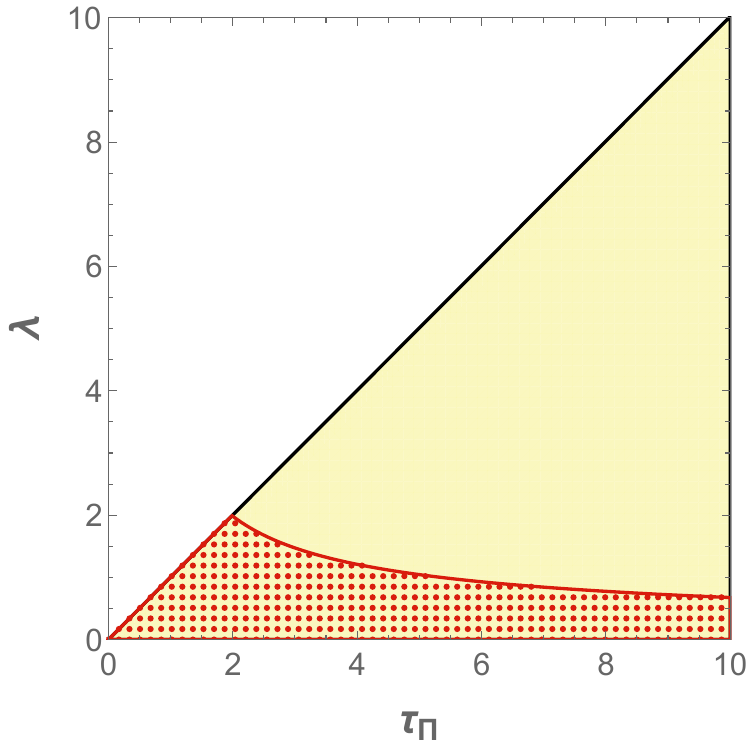}
        \caption{Overlap of \ref{fig:Shear:suffvall} and \ref{fig:Shear:suffv1}}
        \label{fig:Shear:suffoverlap}
    \end{subfigure}
    \caption{Parameter space satisfying ${\rm Im}(\omega)\leq |{\rm Im}(k)|$ for all boosts, and its overlap with $PSS (v\to 1)$ at $O\left(k^6\right)$ with $k = 0.394+ 0.1~i$ in the Shear channel.}
    \label{fig:Shear:sufficient}
\end{figure*}
    \begin{figure*}[ht!]
    \centering
    \begin{subfigure}{0.3\textwidth}
        \centering
        \includegraphics[width=\textwidth]{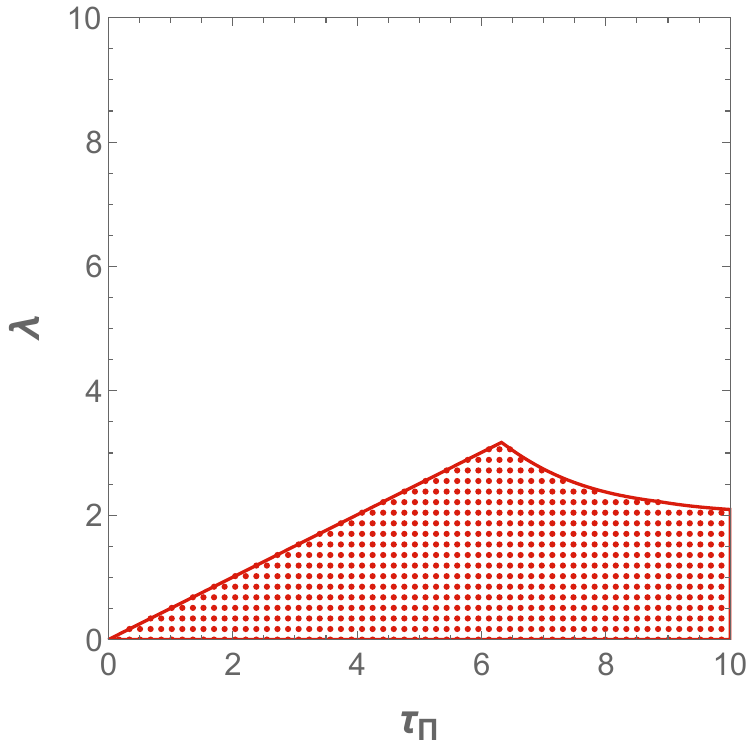}
        \caption{$\bigcap_{v \in (-1,1)} PSS(v)$}
        \label{fig:Sound:suffvall}
    \end{subfigure}
    \hfill
    \begin{subfigure}{0.3\textwidth}
        \centering
        \includegraphics[width=\textwidth]{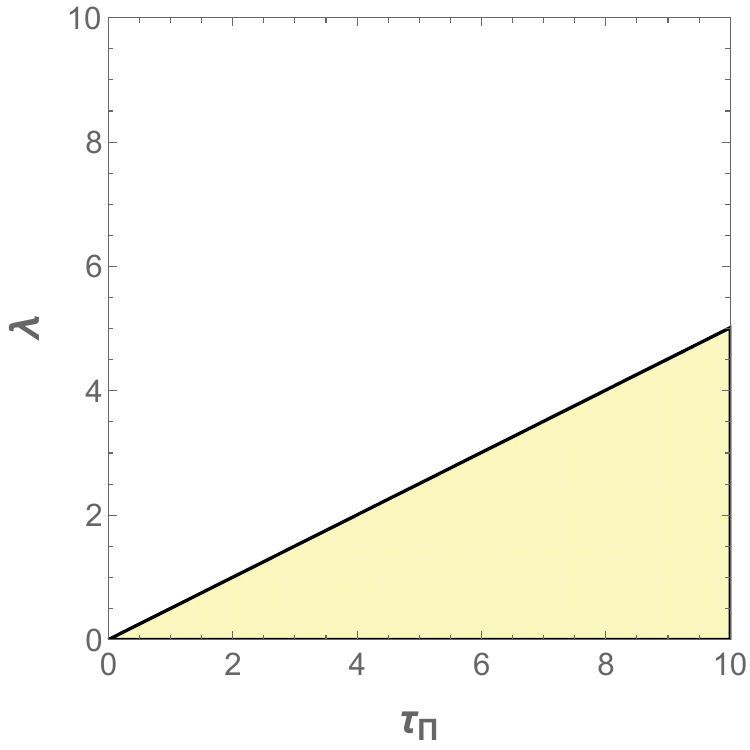}
        \caption{$PSS(v \to 1)$}
        \label{fig:Sound:suffv1}
    \end{subfigure}
    \hfill % Adds horizontal space
    \begin{subfigure}{0.3\textwidth}
        \centering
        \includegraphics[width=\textwidth]{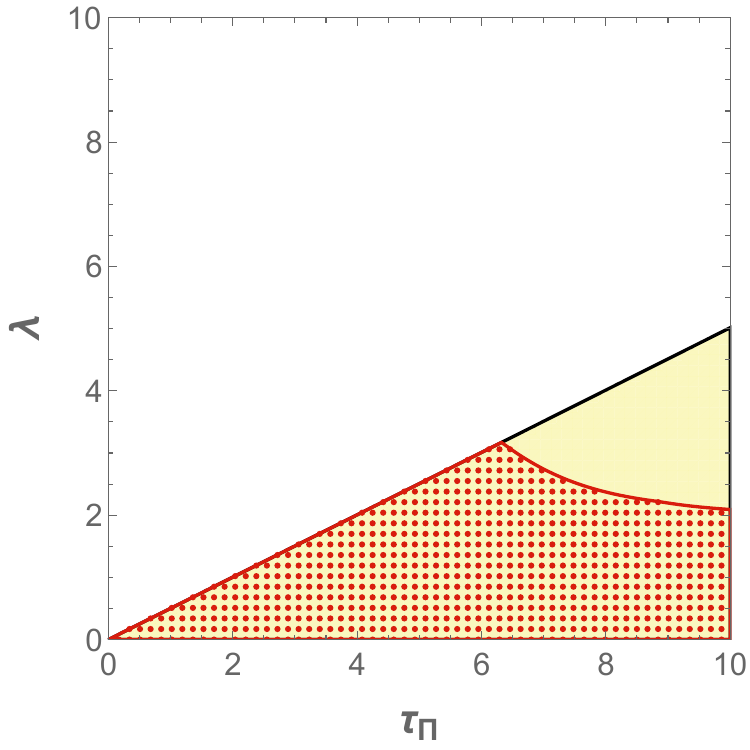}
        \caption{Overlap of \ref{fig:Sound:suffvall} and \ref{fig:Sound:suffv1}}
        \label{fig:Sound:suffoverlap}
    \end{subfigure}
    \caption{Parameter space satisfying ${\rm Im}(\omega)\leq |{\rm Im}(k)|$ for all boosts, and its overlap with $PSS (v\to 1)$ at $O\left(k^6\right)$ with $k = 0.1+ 0.2~i$ in the Sound channel.}
    \label{fig:Sound:sufficient}
\end{figure*}
%\end{widetext}
%\begin{widetext}

In Fig. \ref{fig:Shear:suffvall} and \ref{fig:Sound:suffvall}, we have plotted the region of parameter space where the theory is stable at all velocities taking smaller $v$ intervals \footnote{We have divided the range $v \in(-1,1)$ into 222 equidistant points between $v=-0.999$ and $v=0.999$, and plotted the intersection of $PSS(v)$ for all these values of $v$.} with two representative values of complex-$k$ for MIS shear and sound channels, respectively. For comparison, the $PSS (v\to 1)$ for these cases are also shown in Fig. \ref{fig:Shear:suffv1} and \ref{fig:Sound:suffv1}. In Fig. \ref{fig:Shear:suffoverlap} and \ref{fig:Sound:suffoverlap}, we show the overlap of these two plots, which display the frame-invariant stable parameter space with respect to the $PSS (v\to 1)$ in the respective channels.
Unlike the $k = 0$ situation, where this stability-invariant region was necessary and sufficiently decided by the $PSS (v\to 1)$, for $k\neq0$ it turns out to be a subset of the same. It is bounded by two straight lines, one is the $\tau_\pi$ axis and the other is a fraction of the stability conditions at $v\to 1$, and a curve along which $\lambda$ monotonically decreases with increasing $\tau_\pi$ until $\tau_\pi \to \infty$. This shows that a thin sliver of the parameter space for which the theory is frame-invariantly stable ($PSS(v) ~ \forall~ 0 \leq v <1$) always exists near the $\tau_\pi$-axis for small $\lambda$ values. This can also be shown analytically and presented in Appendix \ref{appen-2}. 

The important point to note is that the frame-invariantly stable parameter space at any $k$ is always enclosed by the $PSS (v\rightarrow 1$). By virtue of $\gamma$-suppression, this $PSS (v\rightarrow 1$) is the same as the $PSS (k=0, v\rightarrow 1$), i.e., the $PSS$ at near-luminal boosts and spatially homogeneous limits (see Fig.~\ref{fig:Shear:k0v1} and \ref{fig:Sound:k0v1}). Thus, our claim holds: the parameter space of stability at near-luminal boosts and the spatially homogeneous limit alone provides the necessary condition for the region corresponding to a frame-invariantly stable, and consequently causal, parameter space at any momentum.

\begin{figure*}[ht!]
    \centering
    \begin{subfigure}{0.3\textwidth}
        \centering
        \includegraphics[width=\textwidth]{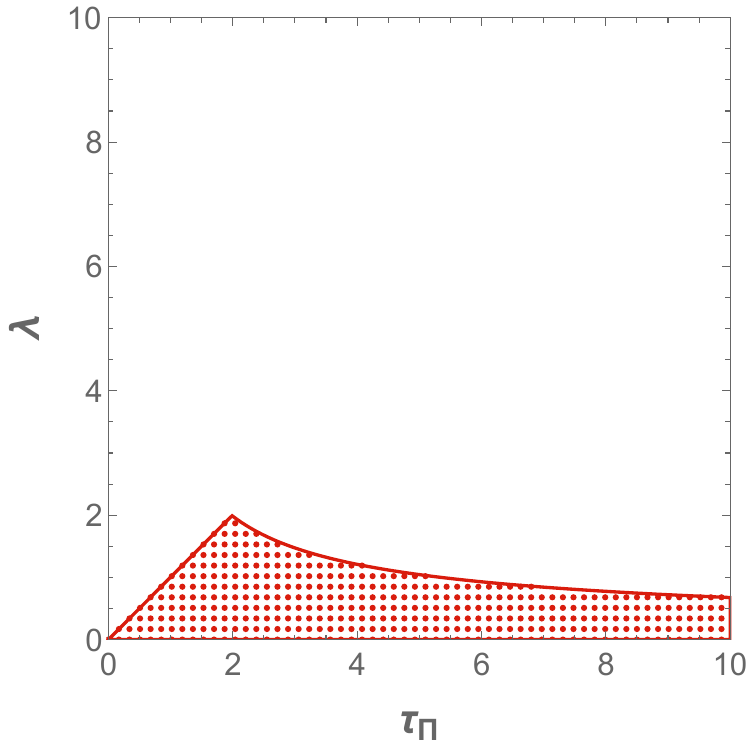}
        \caption{$k=0.394+0.1~i$}
        \label{fig:Shear:nu1}
    \end{subfigure}
    \hfill
    \begin{subfigure}{0.3\textwidth}
        \centering
        \includegraphics[width=\textwidth]{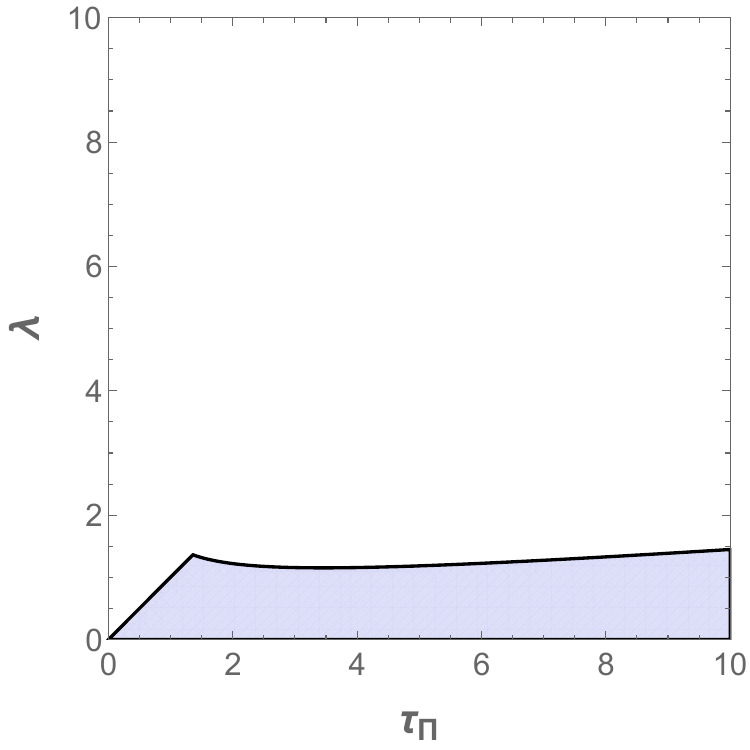}
        \caption{$k=0.57+0.34~i$}
        \label{fig:Shear:nu2}
    \end{subfigure}
    \hfill % Adds horizontal space
    \begin{subfigure}{0.3\textwidth}
        \centering
        \includegraphics[width=\textwidth]{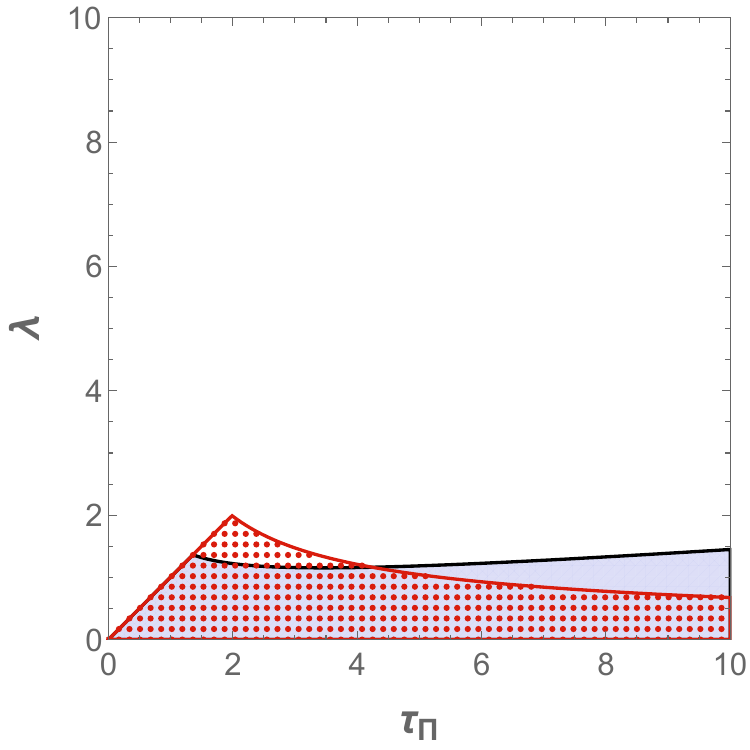}
        \caption{Overlap of \ref{fig:Shear:nu1} and \ref{fig:Shear:nu2}}
        \label{fig:Shear:nu12}
    \end{subfigure}
    \caption{$\bigcap_{v \in (-1,1)} PSS(v)$ for two different values of complex-$k$ and their overlap in the Shear channel}
    \label{fig:Shear:non-unique}
\end{figure*}
    \begin{figure*}[ht!]
    \centering
    \begin{subfigure}{0.3\textwidth}
        \centering
        \includegraphics[width=\textwidth]{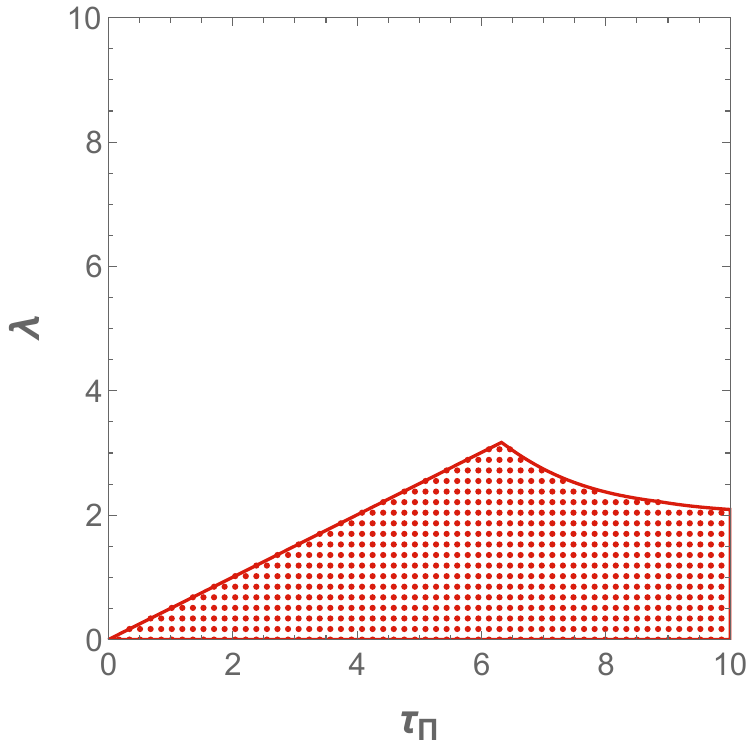}
        \caption{$k=0.1+0.2~i$}
        \label{fig:Sound:nu1}
    \end{subfigure}
    \hfill
    \begin{subfigure}{0.3\textwidth}
        \centering
        \includegraphics[width=\textwidth]{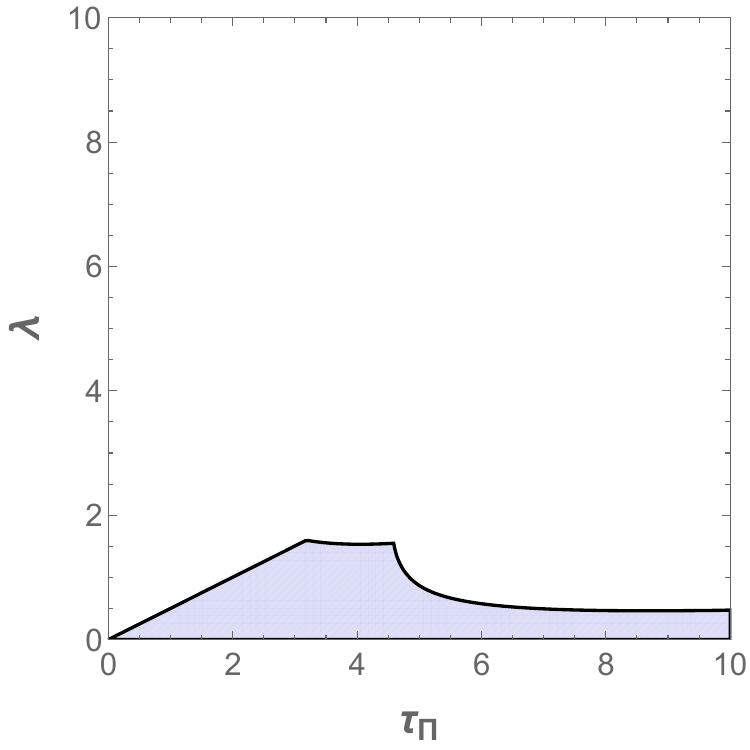}
        \caption{$k=0.3+0.1~i$}
        \label{fig:Sound:nu2}
    \end{subfigure}
    \hfill % Adds horizontal space
    \begin{subfigure}{0.3\textwidth}
        \centering
        \includegraphics[width=\textwidth]{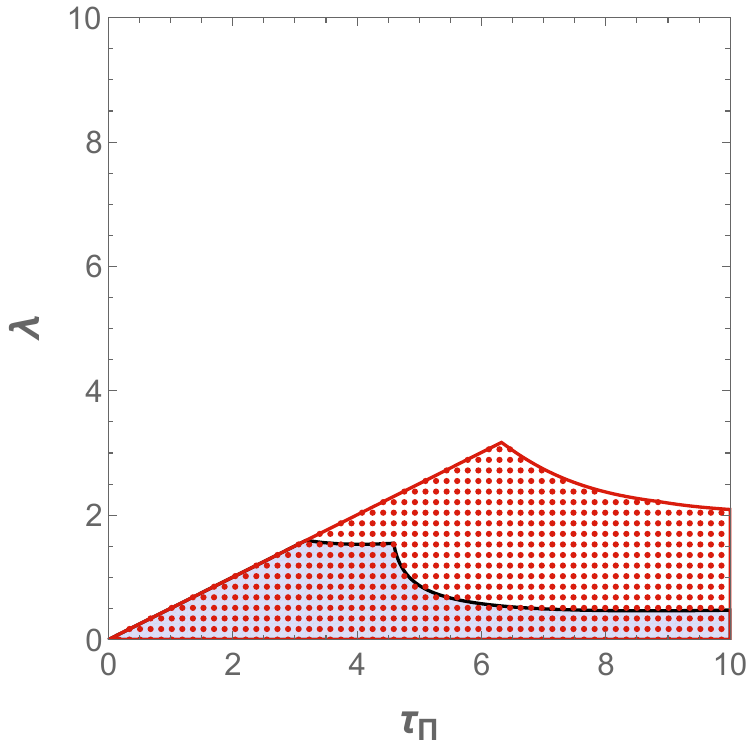}
        \caption{Overlap of \ref{fig:Sound:nu1} and \ref{fig:Sound:nu2}}
        \label{fig:Sound:nu12}
    \end{subfigure}
    \caption{$\bigcap_{v \in (-1,1)} PSS(v)$ for two different values of complex-$k$ and their overlap in the Sound channel}
    \label{fig:Sound:non-unique}
\end{figure*}
%\end{widetext}
%\begin{widetext}
    \begin{figure*}[ht!]
    \centering
    \begin{subfigure}{0.2\textwidth}
        \centering
        \includegraphics[width=\textwidth]{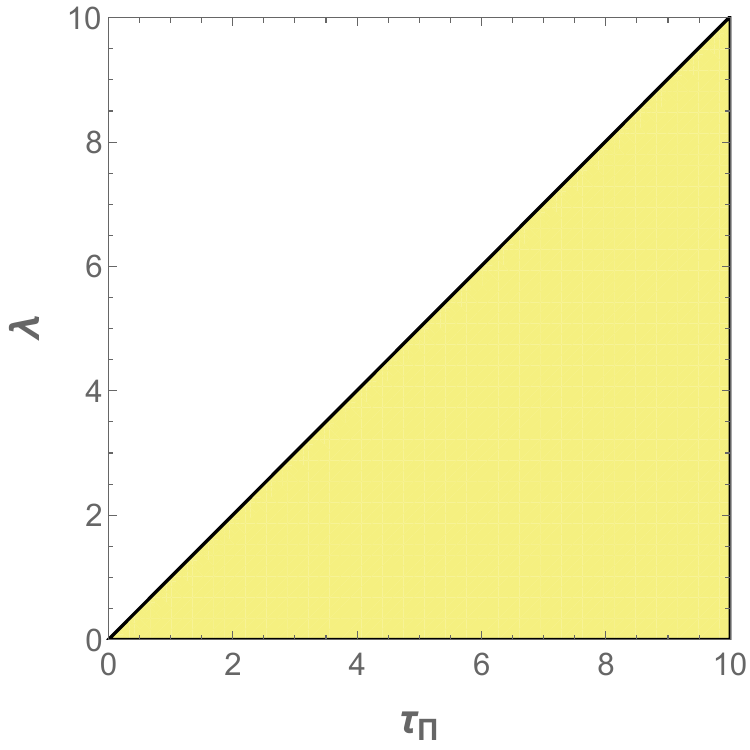}
        \caption{$k=0.45+0.2~i$}
        \label{fig:Shear:$k=0.45+0.2~i$}
    \end{subfigure}
    \hfill % Adds horizontal space
    \begin{subfigure}{0.2\textwidth}
        \centering
        \includegraphics[width=\textwidth]{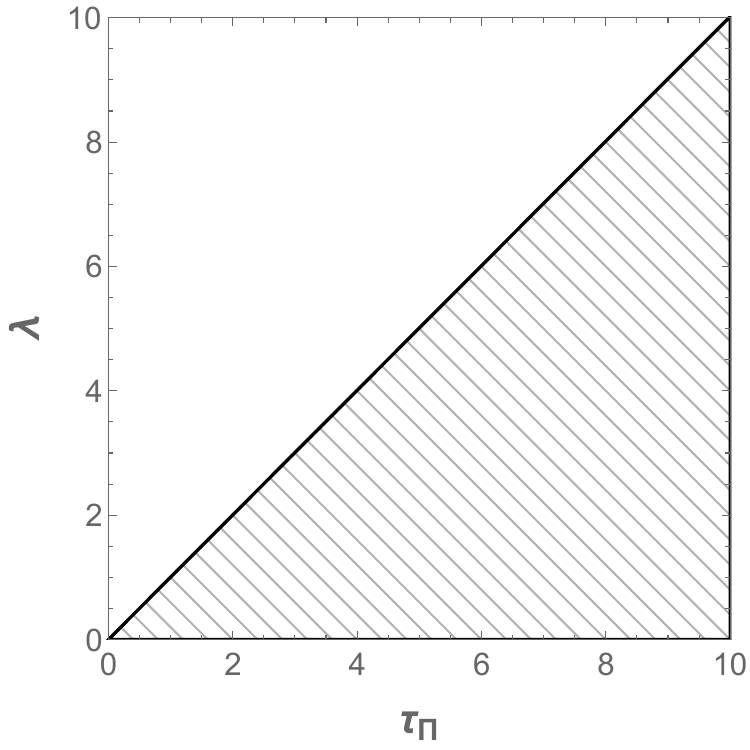}
        \caption{$k=0.12+0.78~i$}
        \label{fig:Shear:$k=0.12+0.78~i$}
    \end{subfigure}
    \hfill
    \begin{subfigure}{0.2\textwidth}
        \centering
        \includegraphics[width=\textwidth]{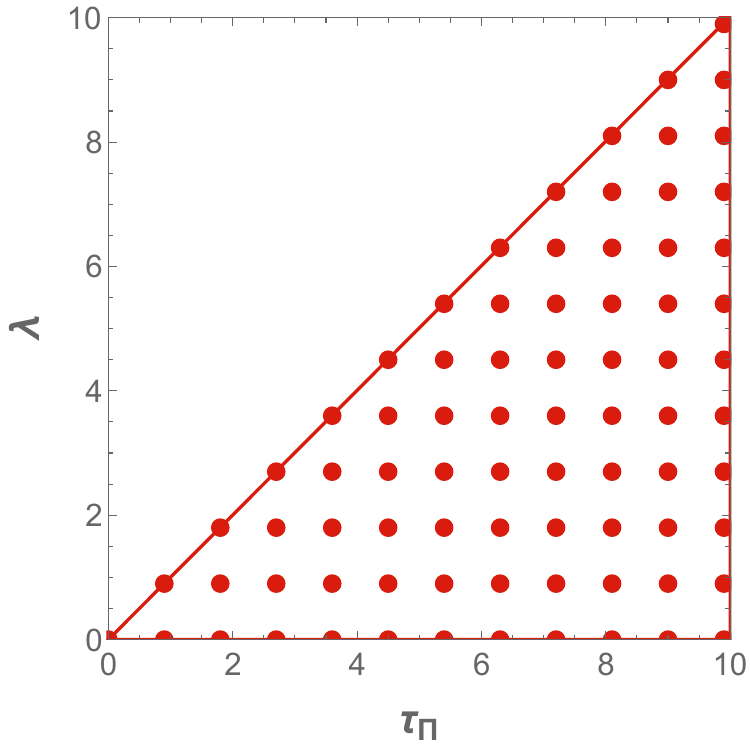}
        \caption{$k=0.85+0.01~i$}
        \label{fig:Shear:$k=0.85+0.01~i$}
    \end{subfigure}
    \hfill
    \begin{subfigure}{0.2\textwidth}
        \centering
        \includegraphics[width=\textwidth]{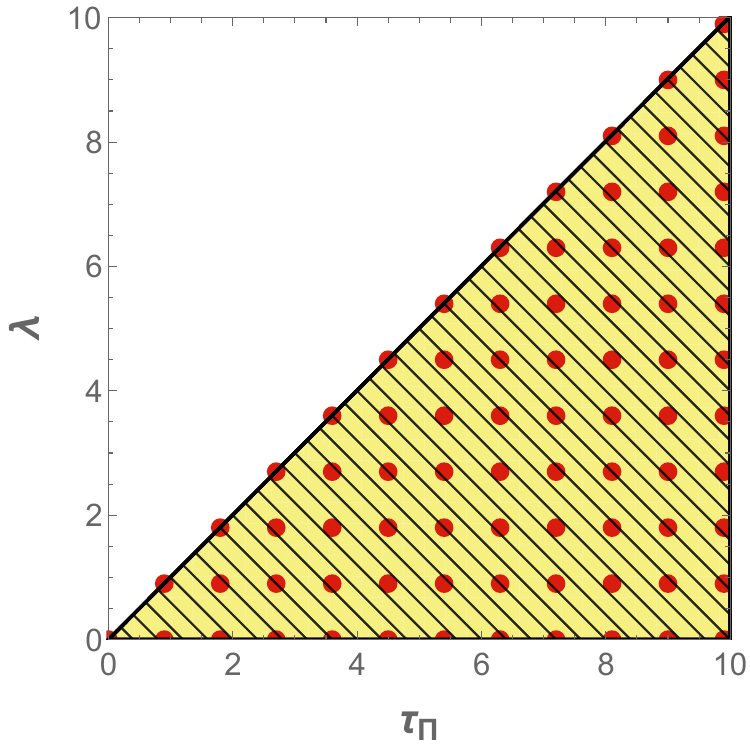}
        \caption{Overlap of \ref{fig:Shear:$k=0.45+0.2~i$},\ref{fig:Shear:$k=0.12+0.78~i$}, \ref{fig:Shear:$k=0.85+0.01~i$}}
        \label{fig:Shear:diffkoverlap}
    \end{subfigure}
    \caption{Parameter space satisfying ${\rm Im}(\omega)\leq |{\rm Im}(k)|$ at $v=0.999$ for different complex values of $k$ in the Shear channel.}
    \label{fig:Shear:diffkv1}
\end{figure*}
    \begin{figure*}[ht!]
    \centering
    \begin{subfigure}{0.2\textwidth}
        \centering
        \includegraphics[width=\textwidth]{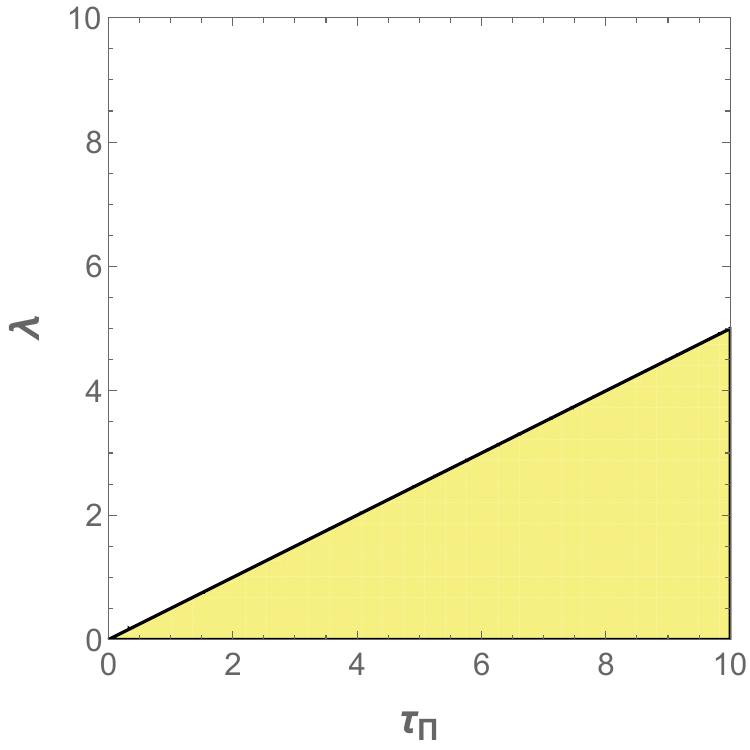}
        \caption{$k=0.45+0.2~i$}
        \label{fig:Sound:$k=0.45+0.2~i$}
    \end{subfigure}
    \hfill % Adds horizontal space
    \begin{subfigure}{0.2\textwidth}
        \centering
        \includegraphics[width=\textwidth]{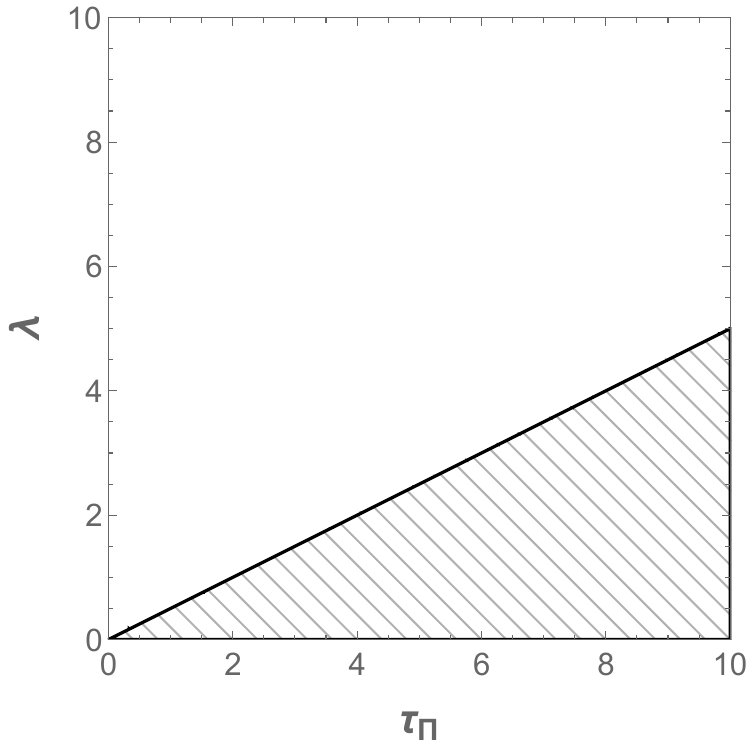}
        \caption{$k=0.12+0.78~i$}
        \label{fig:Sound:$k=0.12+0.78~i$}
    \end{subfigure}
    \hfill
    \begin{subfigure}{0.2\textwidth}
        \centering
        \includegraphics[width=\textwidth]{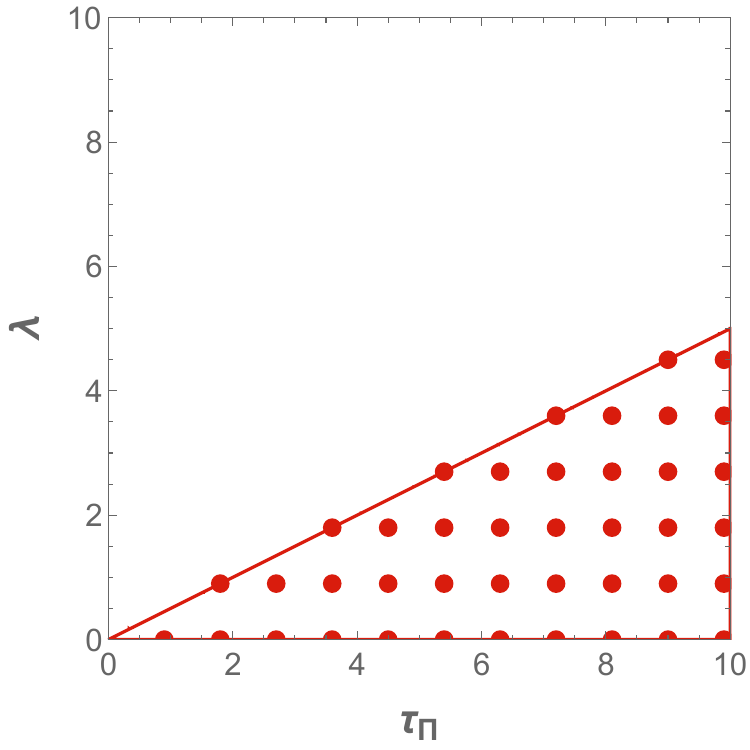}
        \caption{$k=0.85+0.01~i$}
        \label{fig:Sound:$k=0.85+0.01~i$}
    \end{subfigure}
    \hfill
    \begin{subfigure}{0.2\textwidth}
        \centering
        \includegraphics[width=\textwidth]{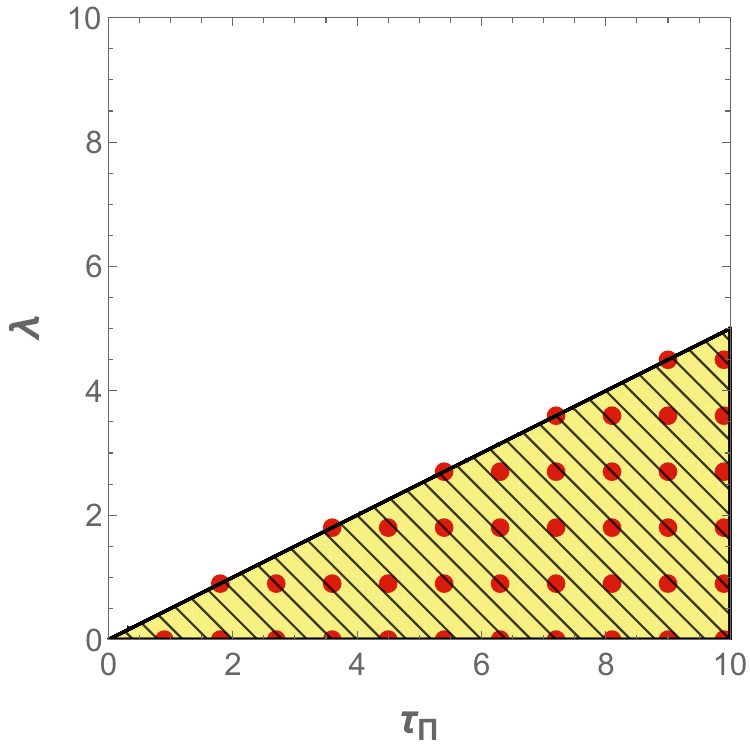}
        \caption{Overlap of \ref{fig:Sound:$k=0.45+0.2~i$},\ref{fig:Sound:$k=0.12+0.78~i$}, \ref{fig:Sound:$k=0.85+0.01~i$}}
        \label{fig:Sound:diffkoverlap}
    \end{subfigure}
    \caption{Parameter space satisfying ${\rm Im}(\omega)\leq |{\rm Im}(k)|$ at $v=0.999$ for different complex values of $k$ in the Sound channel.}
    \label{fig:Sound:diffkv1}
\end{figure*}

However, the $PSC$ satisfying the sufficient condition of causality following the frame-invariant stability condition depends heavily on the value of $k$ used in the analysis. In Figs.~\ref{fig:Shear:non-unique} and \ref{fig:Sound:non-unique}, we plot this region for the MIS shear and sound channels, respectively, with two different values of $k$ for each channel. It can be seen from the plots that although there is some overlap between the parameter space of frame-invariant stability for different values of $k$ (and hence, the sufficient conditions of causality), the regions do not completely enclose or completely overlap with each other. This suggests that using frame-invariant stability criteria to derive sufficient conditions of causality leads to very specific $k$-dependent constraints on the parameter space. These constraints are extremely non-trivial to derive analytically. In contrast, the necessary conditions of causality derived here from $PSS (k=0,v \to 1)$ in the low-wavenumber limit of the theory are analytically tractable and applicable to all values of $k$.

To illustrate this point further, we have included Figs.~\ref{fig:Shear:diffkv1} and \ref{fig:Sound:diffkv1}. We find that the $PSS (v\to 1)$, for different non-zero, complex $k$ values, agree exactly with each other. A comparison with Figs.~\ref{fig:Shear:k0v1} and \ref{fig:Sound:k0v1} reveals that the parameter space is identical to the same at the spatial homogeneous limit ($k=0$) of the theory. This is again an outcome of the $\gamma$-suppression discussed above: as the higher powers of $k/\gamma$ become vanishingly small at $v \to 1$, only the leading non-hydrodynamic term, i.e., the ${\cal{O}}(k^0)$ term in the $\omega$ expansion, contributes to its stability properties.

\begin{figure*}[ht!]
    \centering
    \begin{subfigure}{0.2\textwidth}
        \centering
        \includegraphics[width=\textwidth]{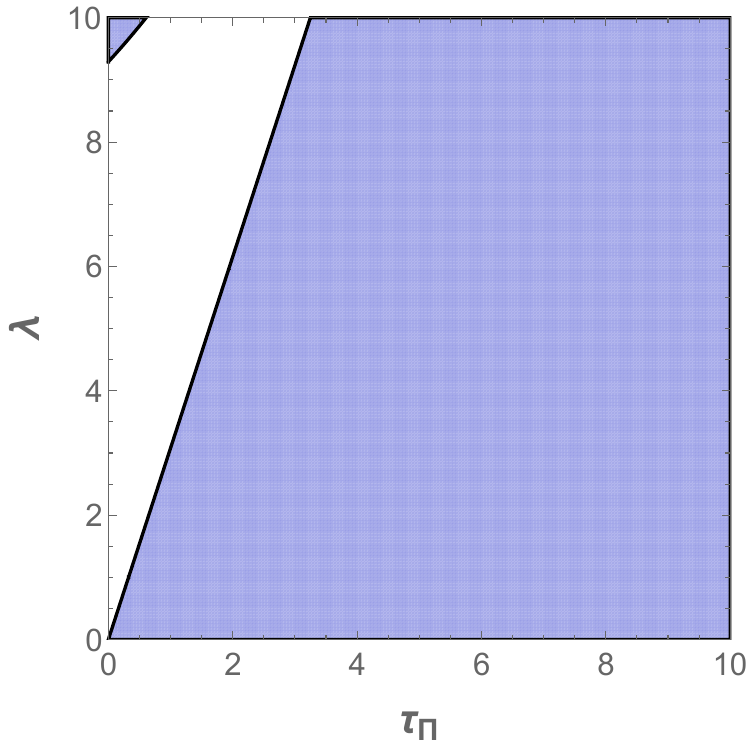}
        \caption{$v=0.57$ up to $O(k^4)$}
        \label{fig:Shear:k4v057}
    \end{subfigure}
    \hfill % Adds horizontal space
    \begin{subfigure}{0.2\textwidth}
        \centering
        \includegraphics[width=\textwidth]{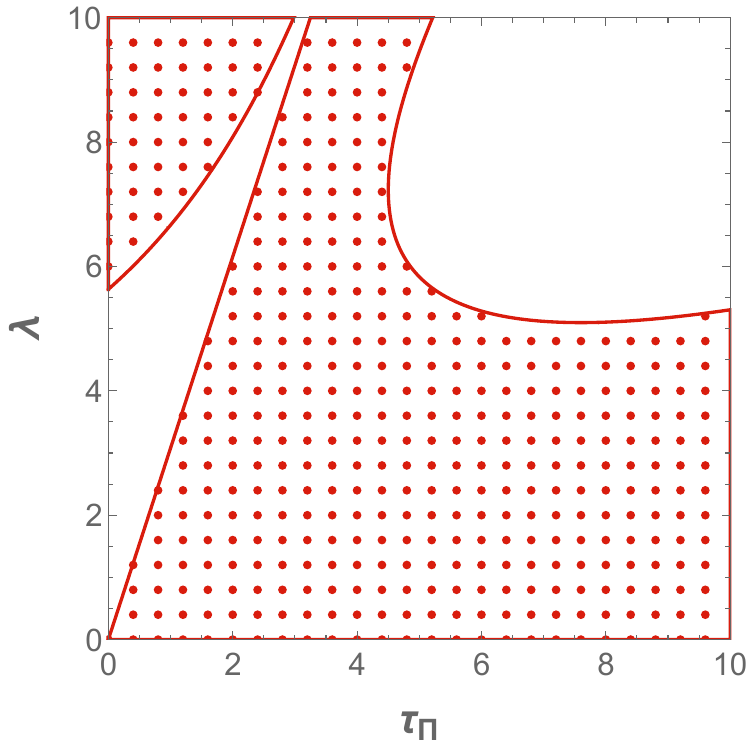}
        \caption{$v=0.57$ up to $O\left(k^6\right)$}
        \label{fig:Shear:k6v057}
    \end{subfigure}
    \hfill
    \begin{subfigure}{0.2\textwidth}
        \centering
        \includegraphics[width=\textwidth]{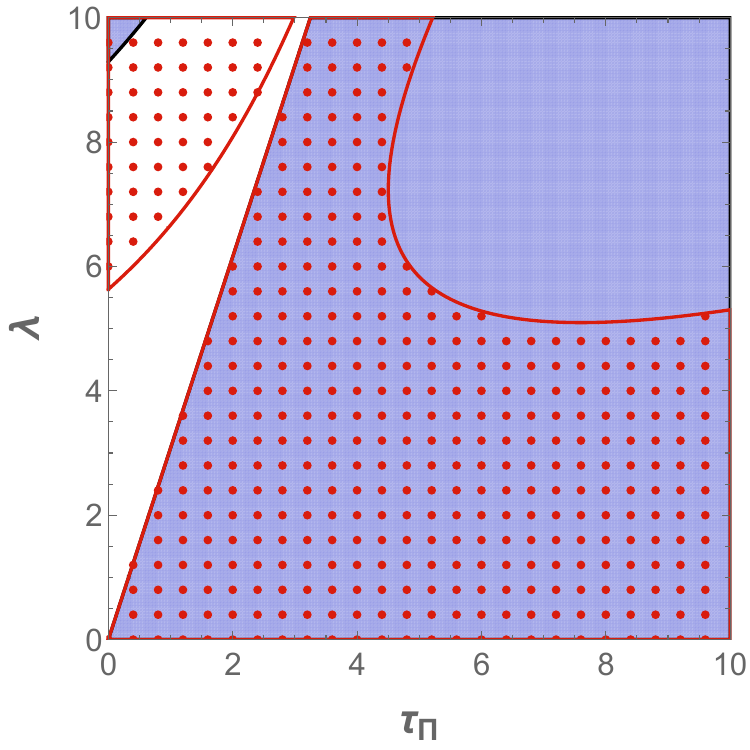}
        \caption{Overlap of \ref{fig:Shear:k4v057} and \ref{fig:Shear:k6v057}}
        \label{fig:Shear:k4k6v057all}
    \end{subfigure}
    \hfill
    \begin{subfigure}{0.2\textwidth}
        \centering
        \includegraphics[width=\textwidth]{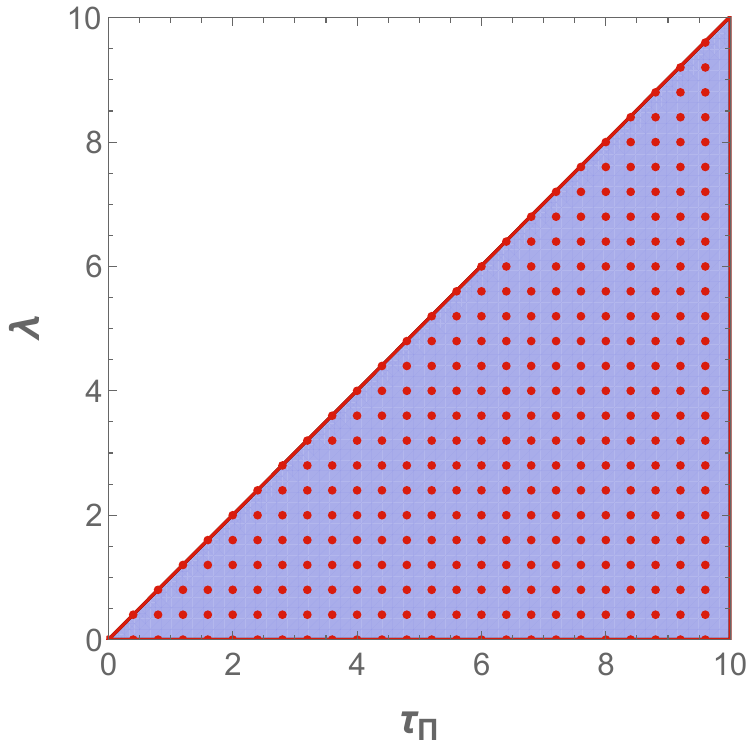}
        \caption{Overlap of $PSS (v\to 1)$ for $O(k^4)$ and $O\left(k^6\right)$}
        \label{fig:Shear:k4k6v1}
    \end{subfigure}
    \caption{Parameter space satisfying ${\rm Im}(\omega)\leq |{\rm Im}(k)|$ for $k=(0.24+0.1~i)$ truncating $\omega$ to different orders in $k$ expansion in the Shear channel.}
    \label{fig:Shear:k4k6full}
\end{figure*}

\begin{figure*}[ht!]
    \centering
    \begin{subfigure}{0.2\textwidth}
        \centering
        \includegraphics[width=\textwidth]{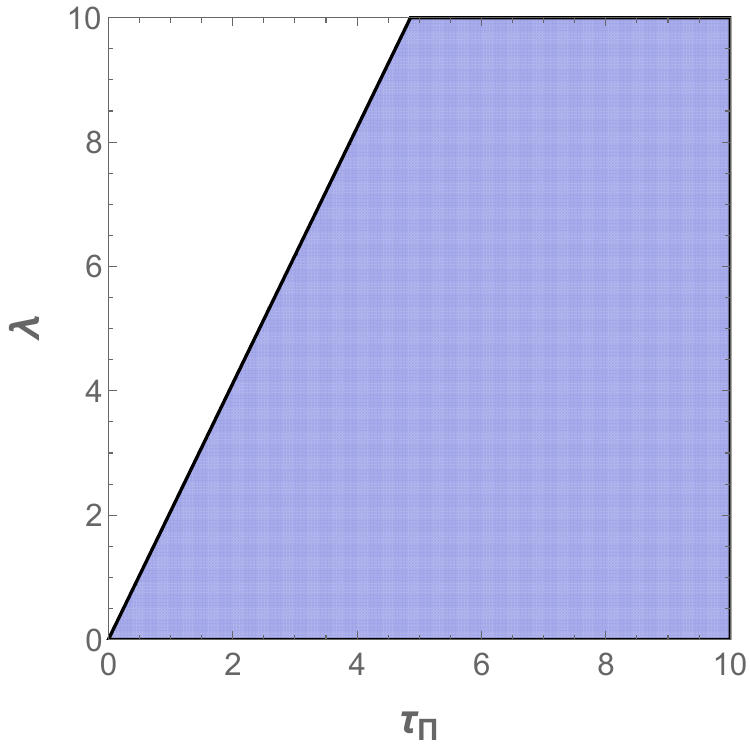}
        \caption{$v=0.57$ up to $O(k^4)$}
        \label{fig:Sound:k4v057}
    \end{subfigure}
    \hfill % Adds horizontal space
    \begin{subfigure}{0.2\textwidth}
        \centering
        \includegraphics[width=\textwidth]{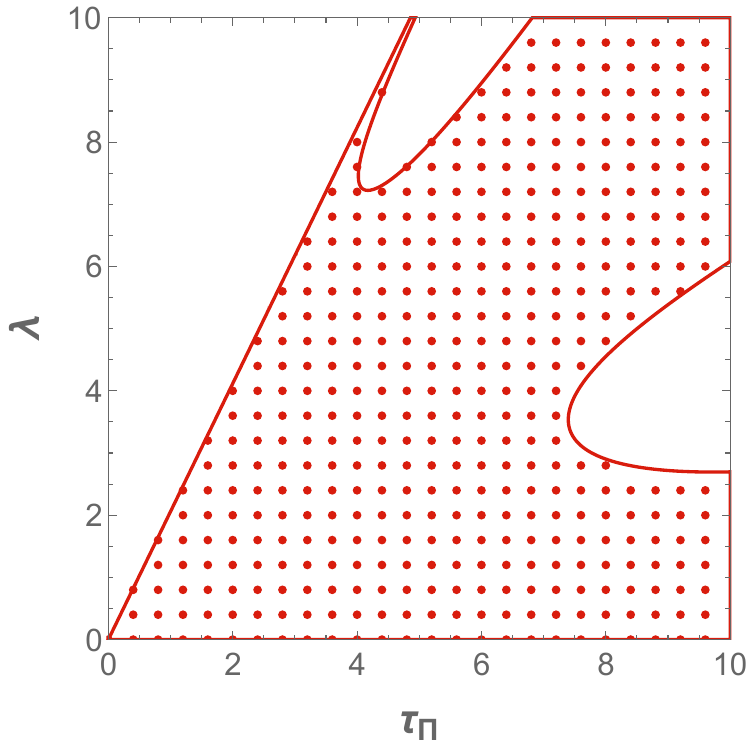}
        \caption{$v=0.57$ up to $O\left(k^6\right)$}
        \label{fig:Sound:k6v057}
    \end{subfigure}
    \hfill
    \begin{subfigure}{0.2\textwidth}
        \centering
        \includegraphics[width=\textwidth]{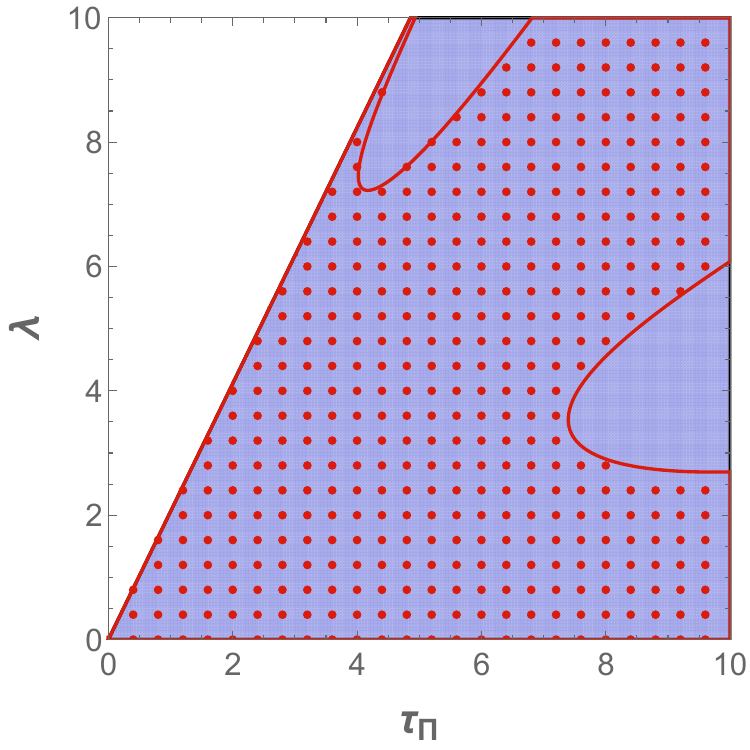}
        \caption{Overlap of \ref{fig:Sound:k4v057} and \ref{fig:Sound:k6v057}}
        \label{fig:Sound:k4k6v057all}
    \end{subfigure}
    \hfill
    \begin{subfigure}{0.2\textwidth}
        \centering
        \includegraphics[width=\textwidth]{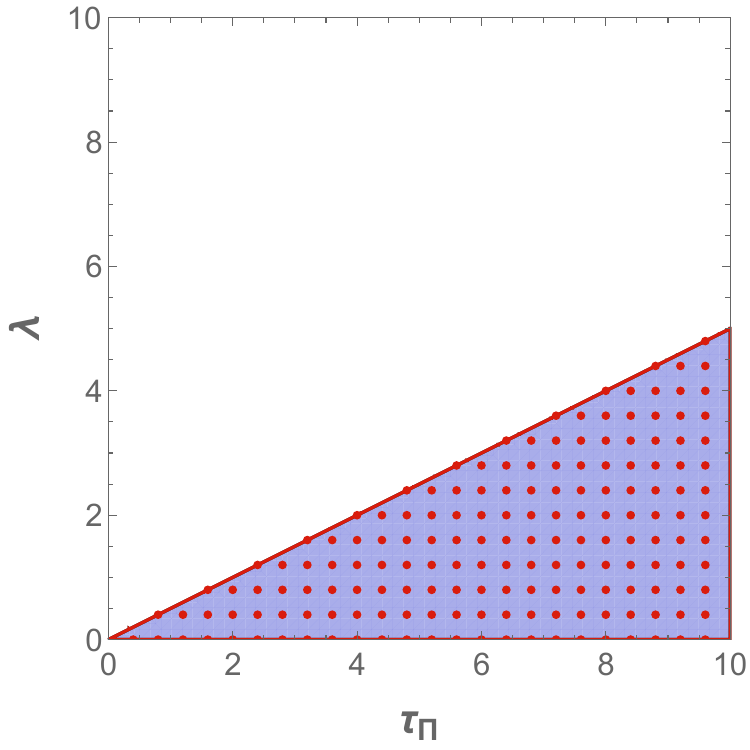}
        \caption{Overlap of $PSS (v\to 1)$ for $O(k^4)$ and $O\left(k^6\right)$}
        \label{fig:Sound:k4k6v1}
    \end{subfigure}
    \caption{Parameter space satisfying ${\rm Im}(\omega)\leq |{\rm Im}(k)|$ for $k=(0.24+0.1~i)$ truncating $\omega$ to different orders in $k$ expansion in the Sound channel.}
    \label{fig:Sound:k4k6full}
\end{figure*}

Another important consequence of $\gamma$-suppression lies in the stability conditions at near-luminal boosts being insensitive to the order of $k$ where the $\omega(k)$ expansion is truncated. At near-luminal velocity, the ``$\gamma$-suppression" is so effective at diminishing the $k$-sensitivity of the stability analysis that including higher and higher orders of $k$-powered terms in $\omega$-expansion does not change the result. We demonstrate this in Figs.~\ref{fig:Shear:k4k6full} and \ref{fig:Sound:k4k6full} for the MIS shear and sound channels respectively, each with two different finite truncations ${{\cal{O}}(k^4)}$ and ${\cal{O}}(k^6)$ of the $\omega(k)$ expansion. 
Figures.~\ref{fig:Shear:k4k6v057all} and \ref{fig:Sound:k4k6v057all} illustrate that for both MIS shear and sound channels, the truncations at $\mathcal{O}(k^4)$ and $\mathcal{O}(k^6)$ for an intermediate boost $v=0.57$ and a representative value of complex-$k=0.24+0.1i$, lead to different $PSS (v)$. As we increase $v$ to $v\to 1$, the $PSS$s coincide identically with each other and those of the spatial homogeneous limit. This is the consequence of the fact that the high-order $(k/\gamma)$ terms in the infinite-series expansion of $\omega$ get suppressed at $v \to 1$, leading to only the $\mathcal{O}(k^0)$ term contributing to the expansion.

% [\textcolor{blue}{I think this is a question that we should answer to complete the full argument and justification of promoting the use of stability analysis at $k\to0, v\to 1$ as a good way of causality analysis.} 
% \textcolor{magenta}{It is indeed a good point. Add in detail with necessary arguments in the conclusion section.}]

% \rs{R.S: see in conclusions.}

At this point, an important question arises: Following this chain of arguments, one could call the $PSS (v)$ of any boost value $v \in (-1,1)$ to correspond to the necessary condition for causality (since the stability-invariant parameter space must be a subspace of any $PSS (v)$ at a given momentum $k$). What makes the $PSS (v\to 1)$ special then? In other words, how unique is the stability condition at $v\to 1$ as a necessary condition for causality? 

To this, we answer the following. Indeed, the stability conditions at any value of $v$ function as necessary conditions for both invariant-stability as well as causality. But as we have shown, it is the $v\rightarrow 1$ case that is insensitive to $k$, and consequently for all $k$, the knowledge of $k\rightarrow0$ stability is all we need in such near-luminal situations. In this regard, the stability conditions at $v \to 1$ are special due to the feature of $\gamma$-suppression.

% Another interesting consequence of $\gamma$-suppression can be seen from Fig. \ref{fig:Shear:k4k6full} and \ref{fig:Sound:k4k6full}. Even if one truncates the $k$-expansion at different finite orders, stability analysis at $v\to 1$ reveals the $PSS$ to be identical for all of these cases.
% Truncations of the $k$-expansion of $\omega$ at different orders,
% Truncating the $k$-expansion of $\omega$ to any arbitrary finite order, if one takes the limit $v \to 1$, 
% Utility, \textcolor{blue}{This also stays stable to different orders in truncation, making it easier to implement numerically.}
% \subsection{$v \to 1$ Results for different $k$'s}
% \subsection{$\mathcal{O}(k^4)$ and $\mathcal{O}(k^6)$ comparison}
% \begin{widetext}
% \end{widetext}
%************************
%************************
\section{Asymptotic Causality from Linearized Stability Analysis?}
\label{sec:gen}
%************************
%************************
Now we are in a position to discuss, from a theoretical perspective, the connection between asymptotic causality and stability analysis at $v\to 1$, which holds even for non-zero complex values of $k$. Through the set of results presented in the previous section, we have claimed that in the MIS theory, linearized stability analysis of the non-hydrodynamic modes at a boost velocity $v\to 1$ leads us to the asymptotic causality criteria of the theory, even at non-zero values of $k$. We argue in this section that our claim holds for a broader class of theories beyond the conformal MIS theory discussed above. This is a generalization of the necessary part of the causality condition, which was derived earlier in~\cite{Roy:2023apk}.

In Appendix A of~\cite{Roy:2023apk}, the authors consider the dispersion polynomial $P_v(\omega,k)$ of a relativistic hydrodynamic theory obeying the following two assumptions:
\begin{enumerate}
    \item \textbf{Assumption 1:} The total power of any term containing $k$ (either of the form $k^n$ or $\omega^m k^n$) must not exceed the maximum power of a pure $\omega^m$ term (i.e., for which $n=0$). In other words,
    \begin{align}
    P_v(\omega,k) = &\sum_{m,n=0}^{M,N} C_{m,n}(v) \omega^m k^n = 0,\nn
    C_{m,n}=0 ~~\forall ~&(m+n ~(\text{where }n\neq 0)) > \nn 
        &\text{max}(m~(\text{where } n=0)).
    \end{align}
    
    \item \textbf{Assumption 2:} In the LRF of the fluid $(v=0)$, the dispersion polynomial $P_v(\omega,k)$ contains only even powers of $k$ making $P_v(\omega,k)$ a scalar. 
\end{enumerate}
The first assumption is a consequence of the causality being satisfied for a given theory, as discussed in~\cite{Bhattacharyya:2025hjs}. The second one restricts the analysis to fluids where there are no other spatial vectors of interest in the LRF, except the wave-vector $\tilde k$. For a theory satisfying these two assumptions, it was argued that the stability conditions at $k\to 0$ yield progressively tighter bounds on the parameter space as the boost increases. Hence, using the argument of frame-invariant stability, this led to the necessary and sufficient condition for causality. This analysis shows that, in a general class of theories beyond MIS or BDNK, the causality criteria can be read off from the linearized stability criteria in the limit $v\to 1, k\to 0$.

However, the crucial point that this argument hinges upon is the monotonic behavior of the $PSS (v)$ with increasing $v$. We observe that this behavior is absent for the case of non-zero momenta ($k \neq 0$). Thus, the chain of arguments given above does not hold anymore to constrain the causal parameter space from linear stability with non-zero wavenumber. 

Instead, we find that we can use the phenomenon of ``$\gamma$-suppression" to argue along similar lines, since it is observed to be a general behavior of relativistic fluid theories in Lorentz-boosted frames~\cite{Bhattacharyya:2025hjs}. Since it exists in any non-spurious modes (modes that are compatible with causality) of any relativistic hydrodynamic theory, the conclusions stemming from this would find relevance in a broader range of theories satisfying the two assumptions mentioned above. Through the following discussion, we argue that reading off the necessary criteria for causality from linearized stability analysis in the limit $(k\to 0, v\to 1)$ serves as a very effective method, due to the generality of the ``$\gamma$-suppression" phenomenon in a general class of theories of relativistic hydrodynamics. Due to this analysis being fully performed in the low-$k$ regime, it remains more aligned with the long-wavelength domain of hydrodynamic validity, as opposed to the large-$k$ asymptotic causality analysis.

The traditional method of deriving the parameter space of a theory, where the theory is stable as well as causal, always comes with added computational complexity. In~\cite{Gavassino:2021owo,Gavassino:2023myj}, it is discussed in detail that in the presence of dissipation, frame-invariant stability is only possible for causal systems. If a dissipative theory is linearly stable in one reference
frame and it is causal, then the theory is stable in all reference frames. So, either we need to test the
stability in all reference frames, which is known to be too cumbersome for a moving fluid, even at $k\rightarrow 0$. Or we can perform asymptotic causality analysis, where using only the LRF ($v=0$) suffices. However, the polynomials in the large-$k$ limit can still be quite difficult to solve.

Now, due to the high complexity of the root-finding algorithms being used in the above methods, it is often quite expensive computationally to derive the roots of two separate polynomials and then unambiguously impose corresponding conditions on them. Methods like R-H analysis or Schur-stability analysis~\cite{Roy:2023apk} do bypass this difficulty, but for more involved dispersion polynomials of higher degree, it often becomes a tedious analysis to impose a long series of inequalities. 
Besides these, it is difficult to generalize the results obtained using these methods without explicitly knowing the parameters (transport coefficients) of the underlying theory or the dependence of the dispersion polynomial's coefficients on them. Thus, while the knowledge of any of these two analyses might give us a more restricted stability-invariant, causal parameter space, the method and the results depend significantly on the boost velocity $v$, the momenta $k$, and the details of the underlying theory. 

Instead of the combination of these two methods, we present here an analytically simpler yet conceptually stronger alternative. We claim that the near-luminal ($v\rightarrow 1$) parameter space at the spatial homogeneous limit ($k\rightarrow 0$) alone can provide the necessarily causal parameter space of the theory at any wavenumber. We have shown that due to the generality of $\gamma$-suppression behavior, stability analysis at $v\to 1$ holds the same results across arbitrary values of $k$, irrespective of the order of truncation in $\omega(k)$ expansion, and suffices to be described by the $k\to 0$ result. 
Now, since the conditions of frame-invariant stability must also include those of the near-luminal boost as well (which for all $k$ gives the same conditions as $k\rightarrow 0$), then, irrespective of the value of $k$, they must be enclosed within the conditions at the $v\to 1, k\to 0$ limit. Hence, the aforementioned stability conditions provide the necessary causality criteria of the theory as well.

This analysis depends only on the leading term of the boosted $\omega(k)$ expansion satisfying the stability criteria. As demonstrated in~\cite {Bhattacharyya:2025hjs} explicitly, this single term in the boosted frame contains the information of all the infinite number of expansion coefficients of the LRF mode $\tilde{\omega}$ in the form of an infinite summation. In contrast, the stability analysis in the rest frame constrains the information of only a finite number of expansion coefficients. So in effect, the derived domain of causality pervades the entire parameter space of any given theory. 

The arguments presented here for the stability constraints in the limit $v\to 1, k\to 0$ serving the purpose of necessary conditions of causality hold true even in more general setups. The analysis presented in Appendix B of~\cite{Roy:2023apk} indicates a connection between linearized stability analysis in the limit $v\to 1, k\to 0$ and asymptotic causality analysis at $k\to \infty$. The stability analysis results at $v \to 1, k \to 0$ remain preserved even for non-zero $k$ (if not sufficiently, at least necessarily). Hence, we conclude that this connection between the stability criteria in the limit $v \to 1, k \to 0$ and the asymptotic causality conditions persists even in a general class of theories, and the necessary condition of causality argued for in the aforementioned discussion with non-zero momenta could lead us to the asymptotic causality condition as well. 

Therefore, with ``$\gamma$-suppression" as the underlying reason and the MIS as a test case presented in the previous sections, we have established that linearized stability analysis at $v \to 1, k \to 0$ serves as an unambiguous check for the necessary condition of causality, while staying completely within the long-wavelength domain of hydrodynamics. Together with the spurious mode method of diagnosing acausal theories in~\cite{Bhattacharyya:2025hjs}, this method provides an effective way of determining whether a theory is causal and of reading off the causality criteria, staying well within the validity of the hydrodynamic theory.
%************************
%************************
\section{Conclusion}
\label{sec:conclu}
%************************
%************************
In this work, we propose an alternative route to derive the necessary conditions of causality in relativistic hydrodynamics solely using linearized stability analysis in Lorentz-boosted frames at non-zero momenta. The central principle is that the boosted dispersion relations exhibit a generic $\gamma$-suppression, such that the expansion of the modes is naturally organized in powers of $k/\gamma$. As the boost approaches the near-luminal regime, $v\rightarrow 1$, the higher-order terms in this expansion become strongly suppressed, and the stability analysis results effectively reduce to those of the leading contribution only, i.e., to the same information that controls the spatially homogeneous limit $k\rightarrow 0$.

We tested this idea explicitly in conformal MIS theory for both the shear and sound channels. At $k = 0$, we recover the results studied in \cite{Roy:2023apk} that the near-luminal stability condition yields the parameter space that is both necessary and sufficient for frame-invariant linear stability, and hence for linearized causality. At non-zero complex momenta, the situation becomes more subtle: the parameter space of stability is no longer monotonic in the boost velocity, and the near-luminal stability region is no longer, by itself, sufficient to characterize the full frame-invariant stable domain. Nevertheless, a key result of the present work is that the frame-invariantly stable region at any non-zero momentum remains enclosed within the near-luminal stability region, which, by $\gamma$-suppression, coincides with the $k\rightarrow 0$ result. Therefore, even at non-zero momenta, the near-luminal stability criterion at the spatially homogeneous limit continues to provide a necessary condition for causality, which agrees with that obtained in the analysis in \cite{Roy:2023apk}.

This also clarifies why the limit $v\rightarrow 1$ is special. In principle, the stability condition at any fixed boost gives a necessary condition for causality, since a causal theory must remain stable in every Lorentz frame and the invariantly stable parameter space must be a subset of the stable parameter space for all boosts. However, the current work shows that generic boosts retain explicit sensitivity to the chosen momentum and the truncation order of the dispersion expansion. The near-luminal limit is distinguished because the feature of $\gamma$-suppression removes this momentum sensitivity: for sufficiently large $\gamma$, the higher-order $k/\gamma$ terms become negligible, so that the stability conditions become effectively independent of the particular non-zero momenta values and of the order at which the series is truncated. This makes the $v\rightarrow 1$ analysis uniquely useful among boosted-frame stability criteria.

From a practical perspective, the method developed here provides a comparatively simple way of reading off the necessary constraints of causality while remaining entirely within the long-wavelength regime of hydrodynamics. This is an effective as well as useful method since the standard asymptotic-causality analysis relies on the large-$k$ limit, which lies outside the strict domain of hydrodynamic validity. At the same time, direct frame-by-frame stability analyses at non-zero momentum are often algebraically and computationally cumbersome. In contrast, the present method reduces the problem to the leading near-luminal stability condition, thereby avoiding much of the complexity associated with solving higher-degree dispersion polynomials.

More broadly, the logic of the argument does not rely on any of the special features of MIS theory alone. Rather, it rests on the general appearance of $\gamma$-suppression in non-spurious boosted modes of relativistic fluid theory and on the relation between causality and frame-invariant stability. For this reason, the present analysis, along with our previous work~\cite{Bhattacharyya:2025hjs}, suggests that the near-luminal linearized stability analysis at $k\rightarrow 0$ can serve as a general and efficient diagnostic of the necessary conditions of causality in a broader class of effective field theories~\cite{Adams:2006sv,deRham:2020zyh,CarrilloGonzalez:2022fwg,Serra:2022pzl}.

Several natural extensions of the current work remain to explore in future. It would be useful to formulate this method directly at the level of general dispersion polynomials, to sharpen the relation between the near-luminal stability criterion and asymptotic causality beyond the MIS example, and to understand more systematically the momentum dependence of the sufficient conditions for causality at non-zero $k$. These questions would further clarify the role of boosted-frame stability as a low-energy probe of relativistic causality.
%************************
%************************
\begin{acknowledgments}
We duly acknowledge Sayantani Bhattacharyya for useful discussions and valuable inputs. R.S. is supported by a postdoctoral fellowship from the West University of Timișoara, Romania.
\end{acknowledgments}
%************************
%************************
\begin{widetext}
\appendix
%************************
%************************
\section{Boosted Dispersion Modes of the MIS Theory up to \texorpdfstring{$\cal O$$ \left(k^6\right)$}{}}
\label{app:allmodesk6}
%************************
%************************
In this appendix, we shall list the expressions of the hydrodynamic and non-hydrodynamic modes of the MIS theory in an arbitrarily boosted frame, following Eq.~\eqref{Lztrans}, up to $\mathcal{O}\left(k/\gamma\right)^6$. Following the conventions developed in Section~\ref{sec:theory}, we express the non-hydrodynamic modes and the hydrodynamic modes, respectively, as 
\ba
\gamma \left( \omega_{nh} + v \,k\right) = \sum_{n=0}^\infty \alpha_n^* \left\{\frac{k}{\gamma}\right\}^n\,, \qquad
\frac{\omega_{h}}{\gamma} = \sum_{n=1}^\infty \tilde a_n \left\{\frac{ k}{\gamma}\right\}^n\,,
\ea
and write down the expressions of $\alpha_n^*$ and $\tilde a_n$ for the different modes of the different channels. The main motivation behind providing these expressions is that we have used the results of the non-hydrodynamic modes up to $\mathcal{O}(k/\gamma)^6$ for the stability-analysis plots in Section~\ref{sec:results}.
%************************
%************************
% \begin{widetext}
\subsection{MIS-Shear: Non-hydro-mode}
%************************
%************************
The non-hydrodynamic coefficients $\alpha_n^*$ for the MIS shear-channel till $n=6$ are:
\begin{equation}
    \begin{split}
        % &\rb{\gamma( \omega_{nh} + v ~k) = \sum_{n=0}^6 \alpha_n^* \left\{\frac{k}{\gamma}\right\}^n + \mathcal{O}\left\{\frac{k}{\gamma}\right\}^7} \\
        \alpha_0^* &= \frac{i}{\lambda  v^2-\tau_\pi }\,, \qquad 
        \alpha_1^* = \frac{2  \lambda  v }{\lambda  v^2-\tau_\pi }\,, \qquad 
        \alpha_2^* = i ~\lambda\,, \qquad
        \alpha_3^* = 2 \lambda ^2 v\,, \qquad
        \alpha_4^* = -i \lambda ^2 \left(5 \lambda  v^2-\tau_\pi \right)\,, \\
        \alpha_5^* &= 2 \lambda ^3 v \left(3 \tau_\pi -7 \lambda  v^2\right)\,, \qquad
        \alpha_6^* = 2 i \lambda ^3 \left(\tau_\pi ^2+21 \lambda ^2 v^4-14 \lambda  \tau_\pi  v^2\right)\,.
    \end{split}
\end{equation}
%************************
%************************
\subsection{MIS-Shear: Hydro-mode}
%************************
%************************
The hydrodynamic coefficients $\tilde{a}_n$ for the MIS shear-channel till $n=6$ are:
\begin{equation}
    \begin{split}
    % &\rb{\frac{\omega_h}{\gamma} = \sum_{n=1}^6 \tilde a_n \left\{\frac{k}{\gamma}\right\}^6 + \mathcal{O}\left\{\frac{k}{\gamma}\right\}^7} \\
    \tilde a_1 &= -v \,, \qquad
    \tilde a_2 = -\frac{i \lambda }{\gamma ^2} \,, \qquad
    \tilde a_3 = -\frac{2 \lambda ^2 v}{\gamma ^2} \,, \qquad
    \tilde a_4 = \frac{i \lambda ^2 \left(5 \lambda  v^2-\tau_\pi \right)}{\gamma ^2} \\
    \tilde a_5 &= \frac{2 \lambda ^3 v \left(7 \lambda  v^2-3 \tau_\pi \right)}{\gamma ^2} \,, \qquad
    \tilde a_6 = -\frac{2 i \lambda ^3 \left(\tau_\pi ^2+21 \lambda ^2 v^4-14 \lambda  \tau_\pi  v^2\right)}{\gamma ^2}\,.
    \end{split}
\end{equation}
%************************
%************************
\subsection{MIS-Sound: Non-hydro-mode}
%************************
%************************
The non-hydrodynamic coefficients $\alpha_n^*$ for the MIS sound-channel till $n=6$ are:

\begin{equation}
    \begin{split}
        % &\rb{\gamma( \omega_{nh} + v ~k) = \sum_{n=0}^6 \alpha_n^* \left\{\frac{k}{\gamma}\right\}^n + \mathcal{O}\left\{\frac{k}{\gamma}\right\}^7} \\
        &\alpha_0^* = -\frac{i \left(v^2-3\right)}{v^2 (4 \lambda +\tau_\pi )-3 \tau_\pi } ,\quad \alpha_1^* = -\frac{24  \lambda  v }{\left(v^2-3\right) \left(v^2 (4 \lambda +\tau_\pi )-3 \tau_\pi \right)} , \quad\alpha_2^* = -\frac{36 i \lambda  \left(v^2+1\right)}{\left(v^2-3\right)^3}, \\
        &\alpha_3^* = -\frac{48 \lambda  v \left(\tau_\pi  \left(v^4-9\right)+2 \lambda  \left(2 v^4+15 v^2+9\right)\right)}{\left(v^2-3\right)^5}, \\
        &\alpha_4^* = \frac{12 i \lambda}{\left(v^2-3\right)^7}  \Big[81 \tau_\pi  (\tau_\pi -4 \lambda )+5 v^8 (4 \lambda +\tau_\pi )^2  + 420 \lambda  v^6 (4 \lambda +\tau_\pi )+126 v^4 (4 \lambda -\tau_\pi ) (10 \lambda +\tau_\pi )\\
        &+108 v^2 \left(20 \lambda ^2-23 \lambda  \tau_\pi +2 \tau_\pi ^2\right)\Big], \\
        &\alpha_5^* = \frac{72 \lambda  v}{\left(v^2-3\right)^9} \Big[\tau_\pi ^3 \left(v^2-3\right)^3 \left(v^2+1\right) \left(v^2+9\right) +12 \lambda  \tau_\pi ^2 \left(v^2-3\right)^2 \left(v^2+3\right) \left(v^4+18 v^2+9\right)\\
        &+64 \lambda ^3 v^2 \left(v^8+43 v^6+324 v^4+567 v^2+189\right) +48 \lambda ^2 \tau_\pi  \left(v^2-3\right) \left(v^8+32 v^6+168 v^4+180 v^2+27\right)\Big], \\
        &\alpha_6^* = -\frac{12 i \lambda}{\left(v^2-3\right)^{11}}  \Big[1792 \lambda ^4 v^4 \big(5346 v^2+\left(v^2+9\right)  \left(v^4+66 v^2+495\right) v^4+1458\big)+16 \lambda  \tau_\pi ^3 \left(v^2-3\right)^3 \times\\
        &\left(918 v^2+7 \left(v^2+6\right) \left(v^2+24\right) v^4+81\right)+\tau_\pi ^4 \left(v^2-3\right)^4 \left(7 \left(v^4+15 v^2+27\right) v^2+27\right)+96 \lambda ^2 \tau_\pi ^2 \left(v^2-3\right)^2 \times \\
        &\left(4860 v^4+2025 v^2+7 \left(v^4+45 v^2+360\right) v^6+81\right) +1792 \lambda ^3 \tau_\pi  v^2 \left(v^2-3\right) \left(v^2+3\right) \left(v^8+57 v^6+504 v^4+513 v^2+81\right)\Big].
    \end{split}
\end{equation}
% \end{widetext}
%************************
%************************
\subsection{MIS-Sound: Hydro-modes}
%************************
%************************
The hydrodynamic coefficients $\tilde{a}_n$ for the MIS sound-channel frequencies $\omega_\pm$ till $n=6$ are:
% \begin{widetext}
\begin{equation}
    \begin{split}
        % \rb{\frac{\omega_\pm}{\gamma}} &= \rb{\sum_{n=1}^6 \tilde a_n \left\{\frac{k}{\gamma}\right\}^6 + \mathcal{O}\left\{\frac{k}{\gamma}\right\}^7} \\
        &\tilde a_1 = \frac{2 v}{v^2-3} \mp \frac{\sqrt{3}}{\gamma ^2 \left(v^2-3\right)} , \quad \tilde a_2 = \frac{18 i \lambda  \left(v^2+1\right)}{\gamma ^2 \left(v^2-3\right)^3} \pm \frac{2 i \sqrt{3} \lambda  v \left(v^2+9\right)}{\gamma ^2 \left(v^2-3\right)^3}, \\
        &\tilde a_3 = \frac{24 \lambda  v}{\gamma ^2 \left(v^2-3\right)^5} \Big[\tau_\pi  \left(v^4-9\right)+2 \lambda  \left(2 v^4+15 v^2+9\right)\Big] \pm \frac{2 \sqrt{3} \lambda}{\gamma ^2 \left(v^2-3\right)^5}  \Big[\tau_\pi  \left(v^2-3\right) \left(v^4+18 v^2+9\right)\\
        &+3 \lambda  \left(v^6+35 v^4+75 v^2+9\right)\Big], \\
        &\tilde a_4 = -\frac{6 i \lambda}{\gamma ^2 \left(v^2-3\right)^7}  \Big[81 \tau_\pi  (\tau_\pi -4 \lambda )+5 v^8 (4 \lambda +\tau_\pi )^2  +420 \lambda  v^6 (4 \lambda +\tau_\pi )+126 v^4 (4 \lambda -\tau_\pi ) (10 \lambda +\tau_\pi )\\
        &+108 v^2 \left(20 \lambda ^2-23 \lambda  \tau_\pi +2 \tau_\pi ^2\right)\Big] \mp \frac{2 i \sqrt{3} \lambda  v}{\gamma ^2 \left(v^2-3\right)^7} \Big[\tau_\pi ^2 \left(v^2-3\right)^2 \left(v^4+30 v^2+45\right)\\
        &+6 \lambda  \tau_\pi  \left(v^2-3\right) \left(v^6+57 v^4+255 v^2+135\right) +10 \lambda ^2 \left(v^8+84 v^6+630 v^4+756 v^2+81\right)\Big], \\
        &\tilde a_5 = \frac{36 \lambda  v}{\gamma ^2 \left(v^2-3\right)^9} \Big[-\tau_\pi ^3 \left(v^2-3\right)^3 \left(v^2+1\right) \left(v^2+9\right) -12 \lambda  \tau_\pi ^2 \left(v^2-3\right)^2 \left(v^2+3\right) \left(v^4+18 v^2+9\right)\\
        &-64 \lambda ^3 v^2 \left(v^8+43 v^6+324 v^4+567 v^2+189\right) -48 \lambda ^2 \tau_\pi  \left(v^2-3\right) \left(v^8+32 v^6+168 v^4+180 v^2+27\right)\Big] \\
        &\pm \frac{2 \sqrt{3} \lambda}{\gamma ^2 \left(v^2-3\right)^9}  \Big[-\tau_\pi ^3 \left(v^2-3\right)^3 \left(v^2+3\right) \left(v^4+42 v^2+9\right)\\
        &+\lambda ^3 \left(-\left(277749 v^4+27702 v^2+7 \left(5 v^6+810 v^4+13527 v^2+48924\right) v^6-729\right)\right)\\
        &-9 \lambda  \tau_\pi ^2 \left(v^2-3\right)^2 \left(v^8+84 v^6+630 v^4+756 v^2+81\right)\\
        &-6 \lambda ^2 \tau_\pi  \left(v^2-3\right) \left(5 v^{10}+615 v^8+7434 v^6+17766 v^4+7857 v^2+243\right)\Big], \\
        &\tilde a_6 = \frac{6 i \lambda }{\gamma ^2 \left(v^2-3\right)^{11}} \Big[1792 \lambda ^4 v^4 \left(5346 v^2+\left(v^2+9\right) \left(v^4+66 v^2+495\right) v^4+1458\right)\\
        &+16 \lambda  \tau_\pi ^3 \left(v^2-3\right)^3 \left(918 v^2+7 \left(v^2+6\right) \left(v^2+24\right) v^4+81\right) +\tau_\pi ^4 \left(v^2-3\right)^4 \left(7 \left(v^4+15 v^2+27\right) v^2+27\right)\\
        &+96 \lambda ^2 \tau_\pi ^2 \left(v^2-3\right)^2 \left(4860 v^4+2025 v^2+7 \left(v^4+45 v^2+360\right) v^6+81\right)\\
        &+1792 \lambda ^3 \tau_\pi  v^2 \left(v^2-3\right) \left(v^2+3\right) \left(v^8+57 v^6+504 v^4+513 v^2+81\right)\Big] \\
        &\pm \frac{2 i \sqrt{3} \lambda  v}{\gamma ^2 \left(v^2-3\right)^{11}} \Big[\tau_\pi ^4 \left(v^2-3\right)^4 \left(v^6+63 v^4+315 v^2+189\right)\\
        &+28 \lambda ^3 \tau_\pi  \left(v^2-3\right) \left(151956 v^6+239355 v^4+89910 v^2+5 \left(v^4+222 v^2+5319\right) v^8+3645\right)\\
        &+12 \lambda  \tau_\pi ^3 \left(v^2-3\right)^3 \left(v^8+116 v^6+1302 v^4+2772 v^2+945\right)\\
        &+12 \lambda ^2 \tau_\pi ^2 \left(v^2-3\right)^2 \left(5 v^{10}+845 v^8+14850 v^6+57834 v^4+53865 v^2+8505\right)\\
        &+126 \lambda ^4 \left(v^{14}+275 v^{12}+8349 v^{10}+63063 v^8+141075 v^6+85833 v^4+7695 v^2-243\right)\Big].
    \end{split}
\end{equation}
% \end{widetext}
%************************
%************************
\section{Proof that a parameter space along positive \texorpdfstring{$\tau_{\pi}$}{} axis (near small \texorpdfstring{$\lambda$}{}) is always stable and causal for all \texorpdfstring{$k$}{} and \texorpdfstring{$v$}{}}
\label{appen-2}
%************************
%************************
A look at the stability plots in Section~\ref{sec:results} reveals that a portion of the parameter space near the positive $\tau_{\pi}$ axis (inside the near-luminal stability triangle) never leaves the stable zone, irrespective of the values of $k$ and $v$. This portion could be a small sliver (meaning $\lambda$ sufficiently small) but it is always there along the positive $\tau_{\pi}$ axis, regardless of any combination of $\{k,v\}$. Below we provide a proof.
%************************
%************************
\subsection{MIS shear channel}
%************************
%************************
For the MIS shear channel, the dispersion relation reads
\begin{align}
\gamma\left(\tau_{\pi}-\lambda^2\right)\omega^2+\left\{i-2\gamma vk\left(\tau_{\pi}-\lambda\right)\right\}\omega  +\left\{-ivk+\gamma k^2\left(\tau_{\pi}v^2-\lambda\right)\right\}=0\,,
\label{MIS-shear}
\end{align}
which can be expressed as
\begin{equation}
 (\omega-vk)\left\{i+\gamma\tau_{\pi}(\omega-vk)\right\}-\gamma\lambda(\omega v-k)^2=0\,.
 \label{shear1}
\end{equation}
At the limit, $\lambda\rightarrow 0$, Eq.~\eqref{shear1} is approximated as
\begin{align}
 (\omega-vk)\left\{i+\gamma\tau_{\pi}(\omega-vk)\right\}=0~.
 \label{shear2}
\end{align}
The first parenthesis of Eq.~\eqref{shear2} provides only hydrodynamic modes with the exact solution
\begin{align}
 \omega=v\,k\,,
\end{align}
whereas the second parenthesis of Eq.~\eqref{shear2} gives only a non-hydrodynamic mode with the exact solution
\begin{align}
 \omega=-\frac{i}{\gamma\tau_{\pi}} + v\,k\,, 
\end{align}
which is stable for all $\{k,v\}$ with the constraint $\tau_{\pi}>0$.
Hence, for sufficiently small $\lambda$, there is always a portion of the $[\lambda~ {\text{vs.}}~\tau_{\pi}]$ parameter space, that is stable and causal for all $k$ and $v$, as long as $\tau_{\pi}$ is positive.
%************************
%************************
\subsection{MIS sound channel}
%************************
%************************
For the MIS sound channel, the dispersion relation reads
\begin{align}
A_3^0\omega^3+\left(A_2^0+A_2^1k\right)\omega^2 +\left(A_1^1k+A_1^2k^2\right)\omega
+\left(A_0^2k^2+A_0^3k^3\right)=0 \,,
 \label{MIS-sound}
\end{align}
where
\begin{align}
 A_3^0&=\tau_{\pi}-\left(\frac{4\lambda}{3}+\frac{\tau_{\pi}}{3}\right){v}^2~, \quad
 A_2^0=i\frac{1}{\gamma}\left(1-\frac{{v}^2}{3}\right)~,\quad
 A_2^1=\left(\frac{8\lambda}{3}-\frac{7\tau_{\pi}}{3}\right){v}+\left(\frac{4\lambda}{3}+\frac{\tau_{\pi}}{3}\right){v}^3~,\quad
 A_1^1=-i\frac{1}{\gamma}\frac{4}{3}{v}~,\nonumber\\
 A_1^2&=-\left(\frac{4\lambda}{3}+\frac{\tau_{\pi}}{3}\right)+\left(\frac{7\tau_{\pi}}{3}-\frac{8\lambda}{3}\right){v}^2~,\quad
 A_0^2=i\frac{1}{\gamma}\left({v}^2-\frac{1}{3}\right)~,\quad
 A_0^3=-\tau_{\pi}{v}^3+\left(\frac{4\lambda}{3}+\frac{\tau_{\pi}}{3}\right){v}~.
\end{align}
One may write Eq.~\eqref{MIS-sound} as
\begin{equation}
 \gamma \tau_{\pi} I_1+\gamma \frac{4}{3}\lambda I_2+iI_3=0~,
\end{equation}
where
\begin{align}
I_1=\left(\omega-vk\right)\left\{(\omega-vk)^2-\frac{1}{3}(\omega v-k)^2\right\},
\quad
I_2=-(\omega-vk)\left(\omega v-k\right)^2 ,
\quad
I_3=(\omega-vk)^2-\frac{1}{3}\left(\omega v-k\right)^2 .
\end{align}
This allows Eq.~\eqref{MIS-sound} to be expressed as
\begin{align}
 \left\{\left(\omega-vk\right)^2-\frac{1}{3}\left(\omega v-k\right)^2\right\}\left[\tau_{\pi}(\omega-vk)+\frac{i}{\gamma}\right] -\frac{4}{3}\lambda(\omega-vk)\left(\omega v-k\right)^2=0~.
 \label{sound1}
\end{align}
In the limit $\lambda\rightarrow 0$, Eq.~\eqref{sound1} reads
\begin{align}
 \left\{\left(\omega-vk\right)^2-\frac{1}{3}\left(\omega v-k\right)^2\right\}\left[\tau_{\pi}(\omega-vk)+\frac{i}{\gamma}\right]=0\,,
 \label{sound2}
\end{align}
where the first parenthesis provides only hydrodynamic modes with the exact solution
\begin{align}
 \omega_{\pm}~=~\left\{\frac{v\pm\frac{1}{\sqrt{3}}}{1\pm\frac{v}{\sqrt{3}}}\right\}~k~, \qquad {\text{causal for all}} ~~\{k,v\}~,
\end{align}
and the second parenthesis gives only non-hydrodynamic mode with the solution
\begin{align}
 \omega=-\frac{i}{\gamma\tau_{\pi}}+vk\,,
 % \quad{\text{stable for all}} ~\{k,v\} ~{\text{as long as}}~\tau_{\pi}>0.
\end{align}
which is stable for all values of $k$ and $v$ with the constraint $\tau_{\pi}>0$.

Thus, for sufficiently small $\lambda$, there is always a region of the $[\lambda~ {\text{vs.}}~\tau_{\pi}]$ parameter space that is stable and causal for all $k$ and $v$, as long as $\tau_{\pi}$ is positive.
\end{widetext}
%==================================================================================================
\bibliographystyle{utphys}
\bibliography{Fin-k}

@article{Bhattacharyya:2024ohn,
    author = "Bhattacharyya, Sayantani and Mitra, Sukanya and Roy, Shuvayu and Singh, Rajeev",
    title = "{Field redefinition and its impact in relativistic hydrodynamics}",
    eprint = "2409.15387",
    archivePrefix = "arXiv",
    primaryClass = "nucl-th",
    doi = "10.1103/PhysRevD.111.014034",
    journal = "Phys. Rev. D",
    volume = "111",
    number = "1",
    pages = "014034",
    year = "2025"
}

@article{Bhattacharyya:2025hjs,
    author = "Bhattacharyya, Sayantani and Mitra, Sukanya and Roy, Shuvayu and Singh, Rajeev",
    title = "{Relativistic Dispersion Spectra across Lorentz boosted frames: Spurious modes and the enigma of causality}",
    eprint = "2512.12450",
    archivePrefix = "arXiv",
    primaryClass = "hep-th",
    month = "12",
    year = "2025",
    journal = ""
}

@article{Roy:2023apk,
    author = "Roy, Shuvayu and Mitra, Sukanya",
    title = "{Causality criteria from stability analysis at ultrahigh boost}",
    eprint = "2306.07564",
    archivePrefix = "arXiv",
    primaryClass = "hep-th",
    doi = "10.1103/PhysRevD.110.016012",
    journal = "Phys. Rev. D",
    volume = "110",
    number = "1",
    pages = "016012",
    year = "2024"
}

@article{Romatschke:2009im,
    author = "Romatschke, Paul",
    title = "{New Developments in Relativistic Viscous Hydrodynamics}",
    eprint = "0902.3663",
    archivePrefix = "arXiv",
    primaryClass = "hep-ph",
    reportNumber = "INT-PUB-09-010",
    doi = "10.1142/S0218301310014613",
    journal = "Int. J. Mod. Phys. E",
    volume = "19",
    pages = "1--53",
    year = "2010"
}

@article{Landau1987Fluid,
    author = "Landau, L. D. and Lifshitz, E. M.",
    title = "Fluid Mechanics: Volume 6 (Course of Theoretical Physics)",    
    journal = "Elsevier Science : Amsterdam",
    year = "1987" 
}

@article{Baier:2007ix,
    author = "Baier, Rudolf and Romatschke, Paul and Son, Dam Thanh and Starinets, Andrei O. and Stephanov, Mikhail A.",
    title = "{Relativistic viscous hydrodynamics, conformal invariance, and holography}",
    eprint = "0712.2451",
    archivePrefix = "arXiv",
    primaryClass = "hep-th",
    reportNumber = "BI-TP-2007-29, INT-PUB-07-45, SHEP-07-47",
    doi = "10.1088/1126-6708/2008/04/100",
    journal = "JHEP",
    volume = "04",
    pages = "100",
    year = "2008"
}

@article{Huovinen:2006jp,
    author = "Huovinen, P. and Ruuskanen, P. V.",
    title = "{Hydrodynamic Models for Heavy Ion Collisions}",
    eprint = "nucl-th/0605008",
    archivePrefix = "arXiv",
    doi = "10.1146/annurev.nucl.54.070103.181236",
    journal = "Ann. Rev. Nucl. Part. Sci.",
    volume = "56",
    pages = "163--206",
    year = "2006"
}

@article{Andersson:2020phh,
    author = "Andersson, Nils and Comer, Gregory L.",
    title = "{Relativistic fluid dynamics: physics for many different scales}",
    eprint = "2008.12069",
    archivePrefix = "arXiv",
    primaryClass = "gr-qc",
    doi = "10.1007/s41114-021-00031-6",
    journal = "Living Rev. Rel.",
    volume = "24",
    number = "1",
    pages = "3",
    year = "2021"
}

@article{Cabral-Rosetti:2001tbx,
    author = "Cabral-Rosetti, Luis G. and Matos, Tonatiuh and Nunez, Dario and Sussman, Roberto A.",
    title = "{Hydrodynamics of galactic dark matter}",
    eprint = "gr-qc/0112044",
    archivePrefix = "arXiv",
    doi = "10.1088/0264-9381/19/14/303",
    journal = "Class. Quant. Grav.",
    volume = "19",
    pages = "3603--3616",
    year = "2002"
}

@article{Hiscock:1983zz,
    author = "Hiscock, W. A. and Lindblom, L.",
    title = "{Stability and causality in dissipative relativistic fluids}",
    doi = "10.1016/0003-4916(83)90288-9",
    journal = "Annals Phys.",
    volume = "151",
    pages = "466--496",
    year = "1983"
}

@article{Hiscock:1985zz,
    author = "Hiscock, William A. and Lindblom, Lee",
    title = "{Generic instabilities in first-order dissipative relativistic fluid theories}",
    doi = "10.1103/PhysRevD.31.725",
    journal = "Phys. Rev. D",
    volume = "31",
    pages = "725--733",
    year = "1985"
}

@article{Hiscock:1987zz,
    author = "Hiscock, William A. and Lindblom, Lee",
    title = "{Linear plane waves in dissipative relativistic fluids}",
    doi = "10.1103/PhysRevD.35.3723",
    journal = "Phys. Rev. D",
    volume = "35",
    pages = "3723--3732",
    year = "1987"
}

@article{Olson:1990rzl,
    author = "Olson, Timothy S.",
    title = "{Stability and causality in the Israel-Stewart energy frame theory}",
    reportNumber = "PRINT-90-0081",
    doi = "10.1016/0003-4916(90)90366-V",
    journal = "Annals Phys.",
    volume = "199",
    pages = "18",
    year = "1990"
}

@article{Kovtun:2012rj,
    author = "Kovtun, Pavel",
    title = "{Lectures on hydrodynamic fluctuations in relativistic theories}",
    eprint = "1205.5040",
    archivePrefix = "arXiv",
    primaryClass = "hep-th",
    doi = "10.1088/1751-8113/45/47/473001",
    journal = "J. Phys. A",
    volume = "45",
    pages = "473001",
    year = "2012"
}

@article{Bemfica:2017wps,
    author = "Bemfica, F{\'a}bio S. and Disconzi, Marcelo M. and Noronha, Jorge",
    title = "{Causality and existence of solutions of relativistic viscous fluid dynamics with gravity}",
    eprint = "1708.06255",
    archivePrefix = "arXiv",
    primaryClass = "gr-qc",
    doi = "10.1103/PhysRevD.98.104064",
    journal = "Phys. Rev. D",
    volume = "98",
    number = "10",
    pages = "104064",
    year = "2018"
}

@article{Denicol:2008ha,
    author = "Denicol, G. S. and Kodama, T. and Koide, T. and Mota, Ph.",
    title = "{Stability and Causality in relativistic dissipative hydrodynamics}",
    eprint = "0807.3120",
    archivePrefix = "arXiv",
    primaryClass = "hep-ph",
    doi = "10.1088/0954-3899/35/11/115102",
    journal = "J. Phys. G",
    volume = "35",
    pages = "115102",
    year = "2008"
}

@article{Pu:2009fj,
    author = "Pu, Shi and Koide, Tomoi and Rischke, Dirk H.",
    title = "{Does stability of relativistic dissipative fluid dynamics imply causality?}",
    eprint = "0907.3906",
    archivePrefix = "arXiv",
    primaryClass = "hep-ph",
    doi = "10.1103/PhysRevD.81.114039",
    journal = "Phys. Rev. D",
    volume = "81",
    pages = "114039",
    year = "2010"
}

@article{Brito:2020nou,
    author = "Brito, C. V. and Denicol, G. S.",
    title = "{Linear stability of Israel-Stewart theory in the presence of net-charge diffusion}",
    eprint = "2007.16141",
    archivePrefix = "arXiv",
    primaryClass = "nucl-th",
    doi = "10.1103/PhysRevD.102.116009",
    journal = "Phys. Rev. D",
    volume = "102",
    number = "11",
    pages = "116009",
    year = "2020"
}

@article{Bemfica:2019knx,
    author = "Bemfica, F{\'a}bio S. and Disconzi, Marcelo M. and Disconzi, Marcelo M. and Noronha, Jorge and Noronha, Jorge",
    title = "{Nonlinear Causality of General First-Order Relativistic Viscous Hydrodynamics}",
    eprint = "1907.12695",
    archivePrefix = "arXiv",
    primaryClass = "gr-qc",
    doi = "10.1103/PhysRevD.100.104020",
    journal = "Phys. Rev. D",
    volume = "100",
    number = "10",
    pages = "104020",
    year = "2019",
    note = "[Erratum: Phys.Rev.D 105, 069902 (2022)]"
}

@article{Bemfica:2020zjp,
    author = "Bemfica, Fabio S. and Disconzi, Marcelo M. and Noronha, Jorge",
    title = "{First-Order General-Relativistic Viscous Fluid Dynamics}",
    eprint = "2009.11388",
    archivePrefix = "arXiv",
    primaryClass = "gr-qc",
    doi = "10.1103/PhysRevX.12.021044",
    journal = "Phys. Rev. X",
    volume = "12",
    number = "2",
    pages = "021044",
    year = "2022"
}

@article{Kovtun:2019hdm,
    author = "Kovtun, Pavel",
    title = "{First-order relativistic hydrodynamics is stable}",
    eprint = "1907.08191",
    archivePrefix = "arXiv",
    primaryClass = "hep-th",
    doi = "10.1007/JHEP10(2019)034",
    journal = "JHEP",
    volume = "10",
    pages = "034",
    year = "2019"
}

@article{Hoult:2020eho,
    author = "Hoult, Raphael E. and Kovtun, Pavel",
    title = "{Stable and causal relativistic Navier-Stokes equations}",
    eprint = "2004.04102",
    archivePrefix = "arXiv",
    primaryClass = "hep-th",
    doi = "10.1007/JHEP06(2020)067",
    journal = "JHEP",
    volume = "06",
    pages = "067",
    year = "2020"
}

@article{Biswas:2022cla,
    author = "Biswas, Rajesh and Mitra, Sukanya and Roy, Victor",
    title = "{Is first-order relativistic hydrodynamics in a general frame stable and causal for arbitrary interactions?}",
    eprint = "2202.08685",
    archivePrefix = "arXiv",
    primaryClass = "nucl-th",
    doi = "10.1103/PhysRevD.106.L011501",
    journal = "Phys. Rev. D",
    volume = "106",
    number = "1",
    pages = "L011501",
    year = "2022"
}

@article{Biswas:2022hiv,
    author = "Biswas, Rajesh and Mitra, Sukanya and Roy, Victor",
    title = "{An expedition to the islands of stability in the first-order causal hydrodynamics}",
    eprint = "2211.11358",
    archivePrefix = "arXiv",
    primaryClass = "nucl-th",
    doi = "10.1016/j.physletb.2023.137725",
    journal = "Phys. Lett. B",
    volume = "838",
    pages = "137725",
    year = "2023"
}

@article{Heller:2022ejw,
    author = "Heller, Michal P. and Serantes, Alexandre and Spali{\'n}ski, Micha{\l} and Withers, Benjamin",
    title = "{Rigorous Bounds on Transport from Causality}",
    eprint = "2212.07434",
    archivePrefix = "arXiv",
    primaryClass = "hep-th",
    doi = "10.1103/PhysRevLett.130.261601",
    journal = "Phys. Rev. Lett.",
    volume = "130",
    number = "26",
    pages = "261601",
    year = "2023"
}

@article{Gavassino:2021owo,
    author = "Gavassino, Lorenzo",
    title = "{Can We Make Sense of Dissipation without Causality?}",
    eprint = "2111.05254",
    archivePrefix = "arXiv",
    primaryClass = "gr-qc",
    doi = "10.1103/PhysRevX.12.041001",
    journal = "Phys. Rev. X",
    volume = "12",
    number = "4",
    pages = "041001",
    year = "2022"
}

@article{Gavassino:2023myj,
    author = "Gavassino, L.",
    title = "{Bounds on transport from hydrodynamic stability}",
    eprint = "2301.06651",
    archivePrefix = "arXiv",
    primaryClass = "hep-th",
    doi = "10.1016/j.physletb.2023.137854",
    journal = "Phys. Lett. B",
    volume = "840",
    pages = "137854",
    year = "2023"
}

@article{Gavassino:2023mad,
    author = "Gavassino, Lorenzo and Disconzi, Marcelo M. and Noronha, Jorge",
    title = "{Dispersion Relations Alone Cannot Guarantee Causality}",
    eprint = "2307.05987",
    archivePrefix = "arXiv",
    primaryClass = "hep-th",
    doi = "10.1103/PhysRevLett.132.162301",
    journal = "Phys. Rev. Lett.",
    volume = "132",
    number = "16",
    pages = "162301",
    year = "2024"
}

@article{Gavassino:2021kjm,
    author = "Gavassino, Lorenzo and Antonelli, Marco and Haskell, Brynmor",
    title = "{Thermodynamic Stability Implies Causality}",
    eprint = "2105.14621",
    archivePrefix = "arXiv",
    primaryClass = "gr-qc",
    doi = "10.1103/PhysRevLett.128.010606",
    journal = "Phys. Rev. Lett.",
    volume = "128",
    number = "1",
    pages = "010606",
    year = "2022"
}

@article{Wang:2023csj,
    author = "Wang, Dong-Lin and Pu, Shi",
    title = "{Stability and causality criteria in linear mode analysis: Stability means causality}",
    eprint = "2309.11708",
    archivePrefix = "arXiv",
    primaryClass = "hep-th",
    doi = "10.1103/PhysRevD.109.L031504",
    journal = "Phys. Rev. D",
    volume = "109",
    number = "3",
    pages = "L031504",
    year = "2024"
}

@article{Hoult:2023clg,
    author = "Hoult, Raphael E. and Kovtun, Pavel",
    title = "{Causality and classical dispersion relations}",
    eprint = "2309.11703",
    archivePrefix = "arXiv",
    primaryClass = "hep-th",
    doi = "10.1103/PhysRevD.109.046018",
    journal = "Phys. Rev. D",
    volume = "109",
    number = "4",
    pages = "046018",
    year = "2024"
}

@article{Fox:1969us,
    author = "Fox, R. and Kuper, C. G. and Lipson, S. G.",
    title = "{Do faster-than-light group velocities imply violation of causality?}",
    doi = "10.1038/223597a0",
    journal = "Nature",
    volume = "223",
    pages = "597",
    year = "1969"
}

@article{Kundt,
    author = "Krotscheck, E. and Kundt, W.",
    title = "Causality criteria",
    journal = " Communications in Mathematical Physics",
    volume = "60",
    pages = "171–180",
    year = "1978"
}

@book{Romatschke:2017ejr,
    author = "Romatschke, Paul and Romatschke, Ulrike",
    title = "{Relativistic Fluid Dynamics In and Out of Equilibrium}",
    eprint = "1712.05815",
    archivePrefix = "arXiv",
    primaryClass = "nucl-th",
    doi = "10.1017/9781108651998",
    isbn = "978-1-108-48368-1, 978-1-108-75002-8",
    publisher = "Cambridge University Press",
    series = "Cambridge Monographs on Mathematical Physics",
    month = "5",
    year = "2019"
}

@article{Heller:2013fn,
    author = "Heller, Michal P. and Janik, Romuald A. and Witaszczyk, Przemyslaw",
    title = "{Hydrodynamic Gradient Expansion in Gauge Theory Plasmas}",
    eprint = "1302.0697",
    archivePrefix = "arXiv",
    primaryClass = "hep-th",
    doi = "10.1103/PhysRevLett.110.211602",
    journal = "Phys. Rev. Lett.",
    volume = "110",
    number = "21",
    pages = "211602",
    year = "2013"
}

@article{Heller:2015dha,
    author = "Heller, Michal P. and Spalinski, Michal",
    title = "{Hydrodynamics Beyond the Gradient Expansion: Resurgence and Resummation}",
    eprint = "1503.07514",
    archivePrefix = "arXiv",
    primaryClass = "hep-th",
    doi = "10.1103/PhysRevLett.115.072501",
    journal = "Phys. Rev. Lett.",
    volume = "115",
    number = "7",
    pages = "072501",
    year = "2015"
}

@article{Israel:1976tn,
    author = "Israel, W.",
    title = "{Nonstationary irreversible thermodynamics: A Causal relativistic theory}",
    doi = "10.1016/0003-4916(76)90064-6",
    journal = "Annals Phys.",
    volume = "100",
    pages = "310--331",
    year = "1976"
}

@article{Israel:1979wp,
    author = "Israel, W. and Stewart, J. M.",
    title = "{Transient relativistic thermodynamics and kinetic theory}",
    doi = "10.1016/0003-4916(79)90130-1",
    journal = "Annals Phys.",
    volume = "118",
    pages = "341--372",
    year = "1979"
}

@article{Muller:1967zza,
    author = "Muller, Ingo",
    title = "{Zum Paradoxon der Warmeleitungstheorie}",
    doi = "10.1007/BF01326412",
    journal = "Z. Phys.",
    volume = "198",
    pages = "329--344",
    year = "1967"
}

@article{Heinz:2013th,
    author = "Heinz, Ulrich and Snellings, Raimond",
    title = "{Collective flow and viscosity in relativistic heavy-ion collisions}",
    eprint = "1301.2826",
    archivePrefix = "arXiv",
    primaryClass = "nucl-th",
    doi = "10.1146/annurev-nucl-102212-170540",
    journal = "Ann. Rev. Nucl. Part. Sci.",
    volume = "63",
    pages = "123--151",
    year = "2013"
}

@article{Gale:2013da,
    author = "Gale, Charles and Jeon, Sangyong and Schenke, Bjoern",
    title = "{Hydrodynamic Modeling of Heavy-Ion Collisions}",
    eprint = "1301.5893",
    archivePrefix = "arXiv",
    primaryClass = "nucl-th",
    doi = "10.1142/S0217751X13400113",
    journal = "Int. J. Mod. Phys. A",
    volume = "28",
    pages = "1340011",
    year = "2013"
}

@article{Braun-Munzinger:2015hba,
    author = {Braun-Munzinger, Peter and Koch, Volker and Sch{\"a}fer, Thomas and Stachel, Johanna},
    title = "{Properties of hot and dense matter from relativistic heavy ion collisions}",
    eprint = "1510.00442",
    archivePrefix = "arXiv",
    primaryClass = "nucl-th",
    doi = "10.1016/j.physrep.2015.12.003",
    journal = "Phys. Rept.",
    volume = "621",
    pages = "76--126",
    year = "2016"
}

@article{Hoult:2024cyx,
    author = "Hoult, Raphael E. and Shukla, Ashish",
    title = "{Causal and stable superfluid hydrodynamics}",
    eprint = "2410.22855",
    archivePrefix = "arXiv",
    primaryClass = "hep-th",
    reportNumber = "CPHT-RR067.092024",
    doi = "10.1007/JHEP04(2025)172",
    journal = "JHEP",
    volume = "04",
    pages = "172",
    year = "2025"
}

@article{Gavassino:2025hwz,
    author = "Gavassino, Lorenzo",
    title = "{Noncovariant parabolic theories of relativistic diffusion}",
    eprint = "2505.18815",
    archivePrefix = "arXiv",
    primaryClass = "gr-qc",
    doi = "10.1103/ppy5-654c",
    journal = "Phys. Rev. D",
    volume = "112",
    number = "3",
    pages = "034026",
    year = "2025"
}

@article{Bhambure:2024axa,
    author = "Bhambure, Jay and Mazeliauskas, Aleksas and Paquet, Jean-Francois and Singh, Rajeev and Singh, Mayank and Teaney, Derek and Zhou, Fabian",
    title = "{Relativistic viscous hydrodynamics in the density frame: Numerical tests and comparisons}",
    eprint = "2412.10303",
    archivePrefix = "arXiv",
    primaryClass = "nucl-th",
    doi = "10.1103/PhysRevC.111.064910",
    journal = "Phys. Rev. C",
    volume = "111",
    number = "6",
    pages = "064910",
    year = "2025"
}

@article{Basar:2024qxd,
    author = {Ba{\c{s}}ar, G{\"o}k{\c{c}}e and Bhambure, Jay and Singh, Rajeev and Teaney, Derek},
    title = "{Stochastic relativistic advection diffusion equation~from the Metropolis algorithm}",
    eprint = "2403.04185",
    archivePrefix = "arXiv",
    primaryClass = "nucl-th",
    doi = "10.1103/PhysRevC.110.044903",
    journal = "Phys. Rev. C",
    volume = "110",
    number = "4",
    pages = "044903",
    year = "2024"
}

@article{Toll:1956cya,
    author = "Toll, John S.",
    title = "{Causality and the Dispersion Relation: Logical Foundations}",
    doi = "10.1103/PhysRev.104.1760",
    journal = "Phys. Rev.",
    volume = "104",
    pages = "1760--1770",
    year = "1956"
}

@article{CarrilloGonzalez:2022fwg,
    author = "Carrillo Gonzalez, Mariana and de Rham, Claudia and Pozsgay, Victor and Tolley, Andrew J.",
    title = "{Causal effective field theories}",
    eprint = "2207.03491",
    archivePrefix = "arXiv",
    primaryClass = "hep-th",
    reportNumber = "Imperial/TP/2022/MC/05",
    doi = "10.1103/PhysRevD.106.105018",
    journal = "Phys. Rev. D",
    volume = "106",
    number = "10",
    pages = "105018",
    year = "2022"
}

@article{Adams:2006sv,
    author = "Adams, Allan and Arkani-Hamed, Nima and Dubovsky, Sergei and Nicolis, Alberto and Rattazzi, Riccardo",
    title = "{Causality, analyticity and an IR obstruction to UV completion}",
    eprint = "hep-th/0602178",
    archivePrefix = "arXiv",
    reportNumber = "CERN-PH-TH-2006-033, HUTP-06-A0005",
    doi = "10.1088/1126-6708/2006/10/014",
    journal = "JHEP",
    volume = "10",
    pages = "014",
    year = "2006"
}

@article{deRham:2020zyh,
    author = "de Rham, Claudia and Tolley, Andrew J.",
    title = "{Causality in curved spacetimes: The speed of light and gravity}",
    eprint = "2007.01847",
    archivePrefix = "arXiv",
    primaryClass = "hep-th",
    reportNumber = "Imperial/TP/2020/CdR/03",
    doi = "10.1103/PhysRevD.102.084048",
    journal = "Phys. Rev. D",
    volume = "102",
    number = "8",
    pages = "084048",
    year = "2020"
}

@article{Serra:2022pzl,
    author = "Serra, Francesco and Serra, Javi and Trincherini, Enrico and Trombetta, Leonardo G.",
    title = "{Causality constraints on black holes beyond GR}",
    eprint = "2205.08551",
    archivePrefix = "arXiv",
    primaryClass = "hep-th",
    doi = "10.1007/JHEP08(2022)157",
    journal = "JHEP",
    volume = "08",
    pages = "157",
    year = "2022"
}
\end{document}